\documentclass[prd,nofootinbib,showpacs,onecolumn]{revtex4}
\usepackage{color}
\usepackage{bm}
\usepackage{graphicx}
\usepackage{amsmath}
\usepackage{amssymb}
\usepackage{enumitem}
\usepackage{hyperref}
\usepackage{subfigure}
\usepackage{array}
\usepackage[english]{babel}
\usepackage{tensor}
\usepackage{xcolor}

\makeatletter
\def\mathcolor#1#{\@mathcolor{#1}}
\def\@mathcolor#1#2#3{%
  \protect\leavevmode
  \begingroup
    \color#1{#2}#3%
  \endgroup
}
\makeatother

\def\be{\begin{equation}}
\def\ee{\end{equation}}
\def\ba{\begin{eqnarray}}
\def\ea{\end{eqnarray}}
\def\l{\left}
\def\r{\right}
\def\f{\frac}

\def\nn{\nonumber}

\allowdisplaybreaks	

\begin{document}

\title{Tackling non-linearities with the effective field theory \\of dark energy and modified gravity}

\author{Noemi Frusciante$^a$ and Georgios Papadomanolakis$^b$ }
\smallskip
\affiliation{$^{a}$ Instituto de Astrof\text{$\acute{i}$}sica e Ci$\hat{e}$ncias do Espa\c{c}o, Departamento de F$\acute{i}$sica da Faculdade
de Ci$\hat{e}$ncias \\ da Universidade de Lisboa, Edif$\acute{i}$cio C8, Campo Grande, P-1749-016 Lisbon,
Portugal\\\smallskip 
$^b$Institute Lorentz, Leiden University, PO Box 9506, Leiden 2300 RA, The Netherlands}



\begin{abstract}
\vspace{0.3cm}
We present the extension of the effective field theory framework to the mildly non-linear scales. The effective field theory approach has been successfully applied to the late time cosmic acceleration phenomenon and it has been shown to be a powerful method to obtain predictions about cosmological observables on linear scales. However, mildly non-linear scales need to be consistently considered when testing gravity theories because a large part of the data comes from those scales. Thus, non-linear corrections to predictions  
on observables coming from the linear analysis can help in discriminating among different gravity theories. We proceed firstly by identifying the  necessary operators which need to be  included in the effective field theory Lagrangian  in order to go beyond the linear order in perturbations and then we construct the corresponding non-linear action. Moreover, we present the complete recipe to map any single field dark energy and modified gravity models into the non-linear effective field theory framework by considering a general action in the Arnowitt-Deser-Misner formalism. In order to illustrate this recipe we proceed to map the beyond-Horndeski theory and low-energy Ho\v rava gravity into  the effective field theory formalism. As a final step we derived the 4th order action in term of the curvature perturbation. This allowed us to identify the non-linear contributions coming from the linear order perturbations which at the next order act like  source terms. Moreover, we confirm that the stability requirements, ensuring the positivity of the kinetic term and the speed of propagation for scalar mode, are automatically satisfied once the viability of the theory is demanded  at linear level. The approach we present here will allow  to construct, in a model independent way, all the relevant predictions on observables at  mildly non-linear scales.
\end{abstract}
\maketitle

\tableofcontents

\section{Introduction}

The cosmological constant problem is challenging standard General Relativity (GR). This led people to propose in place of the cosmological constant new alternatives in the form of  dynamical dark energy (DE) or theories of modified gravity (MG) in order to account for the late time acceleration of the Universe~\cite{Sotiriou:2008rp,Silvestri:2009hh,DeFelice:2010aj,Clifton:2011jh,Tsujikawa:2013fta,Deffayet:2013lga,Joyce:2014kja, Koyama:2015vza,Bull:2015stt}. Collective properties of these alternative proposals are that 1) the dynamics of the graviton turns out to be modified due to the addition of new degrees of freedom (DoFs); 2) the additional DoFs are suppressed on very small scales, as local tests show that GR is very efficient in describing physical phenomena on such scales. Therefore, the GR limit is recovered by demanding the presence of screening mechanisms (see ref.~\cite{Joyce:2014kja} for a review).

It appears clear that we are facing with two scale regimes. One at  large scales, where gravity is modified and standard linear perturbation theory is sufficient to study the growth of small inhomogeneities, and a non-linear regime at small scales,  where screening mechanisms take place hiding any modifications and standard perturbation theory breaks down. However, in this picture, a third regime in between can be also considered which preserves the imprint of a modification of gravity because the screening mechanisms are not fully operating. Indeed, N-body simulations of several theories start exhibiting deviations from the linear results, e.g. in the power spectrum, at scales of $k\gtrsim 0.1$h/Mpc giving an indication of the threshold of validity of  the linear theory\cite{Zhao:2010qy,Barreira:2013eea,Li:2013nua,Winther:2015wla}.  In this intermediate, mildly non-linear regime, one can still employ the standard perturbation theory for modes that are well within the horizon  but one has to go one order further in the perturbative expansion in order to incorporate effects coming from non-linearities~\cite{Maldacena:2002vr,Acquaviva:2002ud,Malik:2003mv,Nakamura:2004rm,Malik:2005cy,Langlois:2006vv,Christopherson:2011hn}.  When one stays beyond the horizon the gradient expansion technique is usually adopted~\cite{Lifshitz:1963ps,Tomita:1975kj,Salopek:1990jq,Tanaka:2006zp,Takamizu:2008ra,Naruko:2012fe,Gumrukcuoglu:2011ef,Takamizu:2013gy,Frusciante:2013haa}.  In the present work we will focus on the mildly non-linear regime and we will go beyond the linear order in perturbation theory as we are interested in sub-horizon modes.

 Going beyond the linear regime from the theory point of view is increasingly becoming a necessity as precision cosmology is probing non-linear scales with high accuracy. Namely, a substantial part of the galaxy clustering, CMB lensing and, most importantly, weak lensing data come from those scales. Usually, signals from intermediate and non-linear scales are cut off from the linear analysis, hence an important amount of information is yet to be accessed. In the intermediate regime, where the deviations from GR are still substantial, gravity theories  leave different imprint on observables, thus new ranges of possibilities to test gravity exist. For instance, modifications of gravity have a strong impact on the clustering of dark matter, then non-linear one loop corrections to the matter power spectrum have be considered~\cite{Koyama:2009me,Bose:2016qun,Bose:2017dtl}.  Furthermore, some MG theories induce additional non-Gaussianities beyond the ones coming from the usual gravitational evolution  which can be studied through the higher order correlation functions, i.e the bispectrum and trispectrum~\cite{Bernardeau:2001qr,Bartolo:2005xa,Bartolo:2013ws}. They offer a possible window into the effects of non-linearities  and can allow  to discriminate between different gravity models. Finally, disentangling DE and MG models is an hard task and the key to distinguish between them potentially lies in the non-linear regime of structure formation. Indeed, MG models due to the presence of a fifth force deeply modify the process of structures formation leaving testable probes  which are difficult to mimic with DE~(see ref.~\cite{Bertschinger:2008zb,Joyce:2016vqv} for a review). 

Testing gravity theories against observations is extremely important yet at the same time difficult because of the large number of models one has to consider. The demand for a unified framework to encompass all single scalar field DE and MG models led to the formulation of the effective field theory approach (hereafter EFT)~\cite{Gubitosi:2012hu,Bloomfield:2012ff,Gleyzes:2013ooa,Bloomfield:2013efa,Piazza:2013coa,Frusciante:2013zop,Gleyzes:2014rba,Gleyzes:2015pma,Perenon:2015sla,Kase:2014cwa,Linder:2015rcz,Frusciante:2016xoj}. This framework has been constructed ad hoc to describe linear order perturbations on large scales by means of a finite number of relevant operators and the resulting action is written in terms of perturbations up to second order.  In this work we will construct, on top of the linear EFT action, a mildly non-linear approach  by adding the necessary operators to the EFT action such that we move to the next leading order in perturbations. In particular, we will construct a 4th order action from which it will be possible to obtain the equations of motion at second order in perturbations.  The resulting extended framework  will allow to have a model independent parametrization of any theory with one extra scalar DoF and at the same time it will preserve a direct link between a specific theory and the EFT language. In this regard, we will provide a general recipe to map a given theory into the non-linear EFT action, thus being of immediate application. The resulting framework will be the theoretical building block to develop all the cosmological observables of interest in follow up works.   

The manuscript is organized as follows. In Sec.~\ref{Sec:EFT}, we present the EFT formalism to describe linear order perturbations, later we extend it beyond the linear order and the new action will come with new operators. In Sec.~\ref{Sec:ADM}, we  present a general Lagrangian for a single scalar field by using the ADM formalism and we describe the general procedure to construct a perturbed action up to a desired order in perturbations. Then, in Sec.~\ref{Sec:restrictedADMaction}, we specialize these action to the class of operators to which  beyond Horndeski and low energy Ho\v rava gravity belong. In Sec.~\ref{Sec:mapping} we translate the EFT action in ADM notation and we work out a general recipe to map any given single scalar field theory in the EFT language. In Secs.~\ref{Sec:bhmapping}-\ref{Sec:horavamapping}, we apply this prescription to the case of beyond Horndeski and low-energy Ho\v rava gravity. In Sec. \ref{Sec:4actionstability}, we derive the action for the non-linear curvature perturbation and discuss the stability of the theory at the next to leading order. Finally, in Sec.~\ref{Sec:conclusion}, we summarize the main results and discuss the potential of the extended EFT framework to mildly non linear regime.


\section{Extending the effective field theory beyond the linear order } \label{Sec:EFT}


An effective approach to describe the linear perturbations of a single scalar field  around a given background has been recently proposed, firstly in the context of Inflation and Quintessence~\cite{Creminelli:2006xe,Cheung:2007st,Weinberg:2008hq,Creminelli:2008wc}, later it has been applied to the late time acceleration with the name EFT of DE/MG~\cite{Gubitosi:2012hu,Bloomfield:2012ff}.  
The EFT of DE/MG  provides an agnostic approach to study linear cosmological perturbations  around a Friedmann-Lemaitre-Robertson-Walker (FLRW) background of  all dark energy and modified gravity models which show an additional scalar DoF. The EFT framework is constructed at the level of the action, in the unitary gauge, and is made up of all the spatial-diffeomorphism invariant operators necessary to describe linear perturbations around a FLRW background. Each individual operator is then accompanied by a time dependent free function, dubbed  EFT function. The choice for the unitary gauge results in the scalar DoF being  absorbed  in the metric, a choice which can be undone by performing the, so called, St\"uckelberg technique, which is realized by restoring the time diffeomorphism invariance by an infinitesimal time coordinate transformation, i.e. $
 t \rightarrow t + \, \pi(x^{\mu})$, where $\pi$ is the explicit DoF.

Let us now consider a flat FLRW  background line element of the form 
\be
ds^2=-dt^2+a(t)^2\delta_{ij}dx^idx^j,
\ee
where $a(t)$ is the scale factor. Then, following the above prescriptions, the quadratic action describing the modified Friedman equations and the linear perturbations around a flat FLRW metric reads 
\begin{align}\label{EFTfirstaction}
\mathcal{S}_{EFT}^{(2)}&=\int d^4x\sqrt{-g}\l[\frac{m_0^2}{2}(1+\Omega(t))R+\Lambda(t)-c(t)\delta g^{00}+\frac{M^4_2(t)}{2}(\delta g^{00})^2-\frac{\bar{M}^3_1(t)}{2}\delta g^{00}\delta K-\frac{\bar{M}^2_2(t)}{2}(\delta K)^2\nonumber\r.\\
&\l.-\frac{\bar{M}_3^2(t)}{2}\delta K^{\mu}_{\nu}\delta K^{\nu}_{\mu}+\f{\hat{M}^2(t)}{2}\delta g^{00}\delta\mathcal{R}+m^2_2(t)h^{\mu\nu}\partial_{\mu}g^{00}\partial_{\nu}g^{00}\r],
\end{align}
where $m_0^2$ is the Planck mass, R and $\mathcal{R}$ are respectively the 4th  and  3rd dimensional Ricci scalar and $K_{\mu\nu}$ the extrinsic curvature associated to the constant time hypersurfaces and K is its trace, $g^{00}$ is the upper time time component of the metric tensor $g^{\mu\nu}$ and $h^{\mu\nu}=g^{\mu\nu}+n^\mu n^\nu$, with $n^\mu$ being the time-like unit vector. Moreover, operators are expanded around the background as  $ A=A_0+\delta A$. Finally, the above action is associated with the usual matter action, $S_m(g_{\mu\nu},\chi_m)$. 

Action~(\ref{EFTfirstaction}) was the first proposal and it includes theories like Horndeski~\cite{Horndeski:1974wa,Deffayet:2009mn}, beyond Horndeski~\cite{Gleyzes:2014dya,Gleyzes:2014qga} and low-energy Ho\v rava gravity~\cite{Horava:2008ih,Horava:2009uw}. Later, it has been generalized  to include a wider class of theories~\cite{Frusciante:2015maa,Frusciante:2016xoj} such as high-energy Ho\v rava gravity~\cite{Blas:2009qj}. We refer the reader to refs.~\cite{Gubitosi:2012hu,Bloomfield:2012ff,Bloomfield:2013efa,Gleyzes:2013ooa,Gleyzes:2014rba,Frusciante:2015maa,Frusciante:2016xoj},  for a complete overview on the mapping of these theories in the EFT framework. In this work we will be mostly interested in beyond Horndeski and low-energy Ho\v rava gravity models, in which case action~(\ref{EFTfirstaction}) is sufficient to describe linear perturbations.

In order to go beyond the linear order in perturbations, one needs to extend the quadratic EFT action. This amount to including all the necessary operators  which contribute to higher order in metric perturbations, as well as expanding the existing operators to the necessary order. At the next to linear order the  equations of motion will be of second order in perturbations, this implies that, at the level of the action, we have to add to the existing action operators which contribute up to 4th-order in perturbations while being negligible at the linear order. We will illustrate with an example why that has to be the case. Let us consider a general field $\phi$ and we will expand it up to the leading orders of interest as 
\be\label{expansion}
\phi=\phi_0+\delta\phi=\phi_0+\delta_1\phi+\f{1}{2}\delta_2\phi+...+\f{1}{n!}\delta_n\phi\,,
\ee
where $\phi_0$ is its homogeneous background value, $\delta \phi$ is its total perturbation part which can be expanded up to an arbitrary  n-th order.  In the present work we are interested in the linear order, i.e. $\delta_1\phi$ and in $\delta_2\phi$ which is  the second order perturbation. In the action we can construct terms of the form $ \sim \delta_1\phi^2\delta_2\phi$, which  is a 4th-order term. When computing the Euler-Lagrange equations  w.r.t. $\delta_1\phi$ we obtain the following term
\be
\f{\partial{L}}{\partial \delta_1\phi}\sim 2\delta_1\phi\delta_2\phi \,\,\rightarrow\,\,\mbox{3rd-order term}\,,
\ee
which can be ignored for the desired equation of motion. On the other hand, when we compute the Euler-Lagrange equation for the $\delta_2\phi$ variable, we get
\be
\f{\partial L}{\partial \delta_2\phi} \sim \delta_1\phi^2 \,\,\rightarrow\,\, \mbox{2nd-order term}\,,
\ee
which will contribute to the equations of motion and hence needs to be considered in the action.  The above argument holds for other combinations as well, such as $\delta_2\phi^2$ or terms  involving derivatives which do not change the perturbative order. Let us note that terms at 4th-order of the form $\delta_1\phi^4$ will be not considered as it is clear from the above  argument that they will not give any contribution to the second order equations of motion.

According to the above arguments many geometrical quantities and their combinations can be constructed. In the following, we will consider only additional operators necessary to deal with theories like Horndeski/beyond Horndeski and low-energy Ho\v rava gravity. The main result is the identification of the following operators as the ones needed to extend the EFT action in order to start including non-linear effects:
\ba\label{new operators}
&&(\delta g^{00})^3\,,\quad (\delta K)^3 \,,\quad (\delta g^{00})^2\delta K\,,\quad \delta g^{00}(\delta K)^2\,,\quad(\delta g^{00})^2\delta \mathcal{R}\,,\quad \delta g^{00}\delta K^\mu_\nu\delta K_\mu^\nu\,, \nn\\
 &&\quad \delta K_\mu^\nu \delta K^\mu_\lambda\delta K_\nu^\lambda\,,\quad \delta K\delta K_\mu^\nu\delta K^\mu_\nu\,\quad \delta g^{00}\delta\mathcal{R}\delta K\, ,\quad\delta g^{00}\delta K^\mu_\nu\delta \mathcal{R}^\nu_\mu \,,\quad h^{\mu\nu}(\partial_{\mu}g^{00}\partial_{\nu}g^{00})\delta g^{00}.
\ea
Note that, as usual, $\delta$ is the total perturbation, which later  will be splitted into the first and second order contribution according to eq.~(\ref{expansion}). They are respectively accompanied by the following new EFT functions\footnote{ The choice of names is based on the following dimensional analysis: we have $[L]=[M^{-1}]$, where L is length and M is mass. Now, looking at the original Lagrangian we can see that $\delta g^{00}$ is dimensionless as expected. So automatically $\mathcal{R}$ and $K$ have the dimension related to the amount of derivatives they possess. So $[\mathcal{R}]=[L^{-2}]=[M^2]$ and $[K]=[L^{-1}]=[M]$.}
\be
M^4_3(t)\,,\quad M_1(t)\,,\quad M^3_1(t)\,,\quad M^2_4(t)\,,\quad M^2_5(t)\,, \quad M^2_6(t)\,,\quad M_2(t)\,,\quad M_3(t)\,,\quad M_4(t)\,,\quad M_5(t)\,, \quad m^2_3(t)\,,
\ee
where the names of the new EFT functions have been chosen according to the mass dimension of the operator. 

So, after identifying the necessary operators, we propose the following action as the one describing the perturbations up to one order beyond the linear one: 
\begin{align}\label{EFTnonlinear}
\mathcal{S}_{EFT}^{(4)}=\int d^4x\sqrt{-g}&\l[\frac{m_0^2}{2}(1+\Omega(t))R+\Lambda(t)-c(t)\delta g^{00} +\frac{M^4_2(t)}{2}(\delta g^{00})^2-\frac{\bar{M}^3_1(t)}{2}\delta g^{00}\delta K-\frac{\bar{M}^2_2(t)}{2}(\delta K)^2\r. \nn\\
&\l.-\frac{\bar{M}_3^2(t)}{2}\delta K^{\mu}_{\nu}\delta K^{\nu}_{\mu}+\f{\hat{M}^2(t)}{2}\delta g^{00}\delta\mathcal{R}+m^2_2(t)h^{\mu\nu}\partial_{\mu}g^{00}\partial_{\nu}g^{00}+M_1(t)(\delta K)^3\r.\nn\\
&\l.+M_4(t)\delta g^{00}\delta\mathcal{R}\delta K+M^2_4(t)\delta g^{00}(\delta K)^2+ M^2_5(t) (\delta g^{00})^2\delta \mathcal{R}+M^3_1(t)(\delta g^{00})^2\delta K \r.\nn\\
&\l.+ M^2_6(t)\delta g^{00}\delta K^\mu_\nu\delta K_\mu^\nu +M^4_3(t)(\delta g^{00})^3+ M_2(t)\delta K_\mu^\nu \delta K^\mu_\lambda\delta K_\nu^\lambda+  M_3(t)\delta K\delta K_\mu^\nu\delta K^\mu_\nu \r.\nn\\
&\l.+M_5(t)\delta g^{00}\delta K^\mu_\nu\delta \mathcal{R}^\nu_\mu+ m^2_3(t) h^{\mu\nu}(\partial_{\mu}g^{00}\partial_{\nu}g^{00})\delta g^{00} \r].
\end{align}

Eq.~(\ref{EFTnonlinear}) represents the extension of the EFT framework to the next to linear order in perturbations. We have identified  11  new operators which need to be included in the original quadratic EFT action in order to study non-linearities for theories belonging to the classes of beyond Horndeski and low-energy Ho\v rava gravity and we have also defined the corresponding EFT functions. 
Let us conclude this Section by noticing that the action presented above does not permit any Ostrogradsky ghosts. This is guaranteed by the building blocks used to construct the new operators  \cite{Gleyzes:2014qga}. In other words, by construction, there will be no higher than second order time derivatives in the equations of motion.

In the next sections we will provide a useful recipe to map a given gravity theory in the EFT framework presented above.


\section{From an ADM action to the effective field theory language}\label{Sec:ADM}


The EFT framework is  powerful in treating model independent parametrization of gravity as well as to study specific theories by mapping them into the EFT formalism. In this section we will present the prescription one needs to follow to map any single scalar field gravity model into the EFT framework.

We first construct a general Lagrangian which includes all the operators one needs in order to parametrize a single scalar field dark energy and modified gravity model up to 4th-order in perturbations. In this regard, we will use the ADM language and we will generalize the prescription in  refs.~\cite{,Kase:2014cwa,Frusciante:2016xoj}. Subsequently  we will rewrite the EFT action~(\ref{EFTnonlinear}) in pure ADM notation. Finally, comparing the two actions we will be able to identify the expressions of the EFT functions in terms of the general Lagrangian. 

In the Arnowitt-Deser-Misner (ADM) formalism~\cite{Gourgoulhon:2007ue} one starts from a 3+1 decomposition described by the following metric
\be \label{ADMmetric}
ds^2=-N^2dt^2+h_{ij}(N^i dt+dx^i)(N^j dt+dx^j)\,,
\ee
where $N(x,t)$ the lapse function, $N^i(x,t)$ the shift vector and $h_{ij}$ the the three dimensional metric describing the equal time hypersurfaces. The normal vector to said hypersurfaces and the corresponding extrinsic curvature look as follows:
\be
n_{\mu}=N\delta_{\mu 0},\qquad K_{\mu\nu}=h^{\lambda}_{\mu}\nabla_{\lambda}n_{\nu}\,.
\ee

We can write down a general Lagrangian which is function of all the geometrical quantities that can be constructed with ADM geometrical objects and give contribution up to 4th order in perturbations. We decided to consider all the operators which allow to include theories with up to to sixth order in spatial derivatives, in this case the Lagrangian can be written as follows~\footnote{Let us note that another operator can be added when writing the general ADM Lagrangian, i.e. the derivative of the lapse function $\dot{N}$. Usually, adding this operator leads to the appearance of more then one extra scalar DoF in the final theory. However, it has been shown that if the initial Lagrangian containing this operator is degenerate, the theory still propagates only one extra scalar DoF~\cite{Langlois:2017mxy}. The inclusion of this operator to the ADM Lagrangian needs to be done carefully. We will leave for future work the investigation of such additional operator when will will also consider an even more general EFT action.}:
\be\label{general}
L(t,N,K,\mathcal{S},\mathcal{R},\mathcal{U},\mathcal{Z},\mathcal{Z}_1,\mathcal{Z}_2,\alpha_1,\alpha_2,\alpha_3,\alpha_4,\alpha_5,\tilde{\mathcal{K}},\tilde{\mathcal{K}}_1,\tilde{Z},\tilde{Z}_1,\alpha_6,\alpha_7,\alpha_8,\alpha_9)\,,
\ee
where specifically the operators are 
\ba\label{ADM_quantities}
&&\mathcal{S}=K_{\mu\nu}K^{\mu\nu}\,,\,\,{\mathcal Z}={\mathcal R}_{\mu\nu}{\mathcal R}^{\mu\nu}\,,\,\,{\mathcal U}={\mathcal R}_{\mu\nu}K^{\mu\nu}\,,\,\,{\mathcal Z}_1=\nabla_i{\mathcal R}\nabla^i{\mathcal R}\,,\,\,{\mathcal Z}_2=\nabla_i{\mathcal R}_{jk}\nabla^i{\mathcal R}^{jk}\,,\nonumber\\
&&\tilde{\mathcal{K}}=K_{ij}K^{ik}K^{j}_k\,, \,\,  \tilde{\mathcal{K}}_1=K_{ij}\mathcal{R}^{ik}K^{j}_k  ,\,\,\tilde{Z}=\mathcal{R}_{ij}\mathcal{R}^{ik}\mathcal{R}^{j}_k, \,\,\tilde{Z}_1=K_{ij}\mathcal{R}^{ik}\mathcal{R}^{j}_k,\nn\\
&&\,\,\alpha_1=a^ia_i\,,\,\,\alpha_2=a^i\Delta a_i\,,\,\,\alpha_3={\mathcal R}\nabla_ia^i\,,\,\,\alpha_4=a_i\Delta^2a^i\,,\,\,\alpha_5=\Delta {\mathcal R}\nabla_ia^i,\nn\\
&&\alpha_6=\mathcal{R}_{ij}a^ia^j\,, \,\,\alpha_7=K_{ij}a^ia^j\,, \,\,\alpha_8=\Delta\mathcal{R}_{ij}a^ia^j\,,\,\,\alpha_9=\Delta K_{ij}a^ia^j,
\ea
with $\Delta=\nabla_k\nabla^k$ and $a^i$  is the acceleration of the normal vector, $n^{\mu}\nabla_{\mu}n_{\nu}$. $\nabla_\mu$ and $\nabla_k$ are  the covariant derivatives constructed respectively with the four dimensional metric, $g_{\mu\nu}$ and the three metric, $h_{ij}$. 

Now, we can  construct the action beyond the linear order as follows
\be
S^{(2)}_{ADM}=\int{}d^4x\sqrt{-g}L(t,N,K,\mathcal{S},\mathcal{R},\mathcal{U},\mathcal{Z},\mathcal{Z}_1,\mathcal{Z}_2,\alpha_1,\alpha_2,\alpha_3,\alpha_4,\alpha_5,\tilde{\mathcal{K}},\tilde{\mathcal{K}}_1,\tilde{Z},\tilde{Z}_1,\alpha_6,\alpha_7,\alpha_8,\alpha_9) \,,
\ee
where the Lagrangian ~(\ref{general}) has been expanded according to:
\be\label{Lagrangianexpansion}
L=\sum_{m=0}^3\sum_{n_1+...+n_d=m}\f{(\delta N)^{n_1}...(\delta \alpha_9)^{n_d}}{n_1!...n_d!}\l(\f{\partial^m L}{\partial \delta N^{n_1}...\partial \delta\alpha_9^{n_d}}\r)_{\delta N,..., \delta \alpha_9=0}+\mathcal{O}(4)\,.
\ee
Let us note that stopping at 3rd order in derivatives in the above expansion means that we are considering 3rd order terms in the total perturbation, i.e. $\delta^3$. However, when we expand $\delta$ in its linear and second order perturbations following eq.~(\ref{expansion}), the $\delta^3$ terms will give perturbations up to 4th-order, such as $\delta_1^2 \delta_2$ accordingly to what we have demanded.

Once that has been done, we finally obtain the non-linear action in the ADM formalism up till the desired order. The prescription described so far is very general and can be applied to any set of operators. In the upcoming subsections we will make this approach explicit for a subset of operators and provide the final mapping results.


\subsection{Non-linear action in ADM formalism}\label{Sec:restrictedADMaction}


In this section we will apply the procedure described in  the previous section to the subset of operators to which beyond Horndeski and low-energy Ho\v rava gravity belong. Therefore, we will restrict the form of the Lagrangian~(\ref{general}) to the following one
\be\label{lagrangianrestricted}
L\equiv L(t,N,K,\mathcal{S},\mathcal{U}, \tilde{\mathcal{K}},\mathcal{R}, \alpha_1)\,, 
\ee
and the  operators will be decomposed according to eq.~(\ref{expansion}) up to 4th-order as follows:
\ba
&&N=1+\delta N=1+\delta_1 N+\f{1}{2}\delta_2N\,, \qquad K=-3H+\delta K=-3H+\delta_1K+\f{1}{2}\delta_2K\,,\quad \alpha_1=\delta \alpha_1 =\delta_1\alpha_1+\f{1}{2}\delta_2\alpha_1\,, \nn\\
&& K^\nu_\mu=-H\delta^\nu_\mu+\delta K^\nu_\mu=-H\delta^\nu_\mu+\delta_1 K^\nu_\mu+\f{1}{2}\delta_2 K^\nu_\mu \,, \quad \mathcal{R}=\delta \mathcal{R}=\delta_1 \mathcal{R}+\f{1}{2}\delta_2 \mathcal{R}\,, \quad
\mathcal{R}^\nu_\mu=\delta_1 \mathcal{R}^\nu_\mu+\f{1}{2}\delta_2 \mathcal{R}^\nu_\mu \,,
\nn\\
&&\mathcal{S}=K^\nu_\mu K^\mu_\nu= 3H^2+\delta \mathcal{S}= 3H^2-2H\delta K+\delta K^\nu_\mu \delta K^\mu_\nu\nn\\
&&\hspace{0.3cm}=3H^2-2H\delta_1 K-H\delta_2 K+ \delta_1 K^\nu_\mu \delta_1 K^\mu_\nu+ \delta_1 K^\nu_\mu \delta_2 K^\mu_\nu+\f{1}{4} \delta_2 K^\nu_\mu \delta_2 K^\mu_\nu \,,\nn\\
&&\mathcal{U}=\mathcal{R}^\mu_\nu K_\nu^\mu=\delta \mathcal{U}=-H\delta_1\mathcal{R}-\f{1}{2}H\delta_2 \mathcal{R}+\delta_1 \mathcal{R}^\mu_\nu\delta_1K^\nu_\mu+\f{1}{2}\delta_1 \mathcal{R}^\mu_\nu\delta_2K^\nu_\mu+\f{1}{2}\delta_1 K^\nu_\mu\delta_2\mathcal{R}^\mu_\nu+\f{1}{4}\delta_2 \mathcal{R}^\nu_\mu\delta_2 K^\mu_\nu \,,\nn\\
&&\tilde{\mathcal{K}}=-3H^3+\delta\tilde{\mathcal{K}}=-3H^3-3H\delta\mathcal{S}+3H^2\delta K +\delta_1K^l_j\delta_1K^k_l\delta_1K^j_k+\f{3}{2}\delta_2K^l_j\delta_1K^k_l\delta_1K^j_k\,, 
\ea
where $H=\dot a/a$ is the Hubble function. Let us note that $\delta_1$ contains linear order terms and might have also second order terms obtained by combinations of the first order metric perturbations. Then, with $\delta_2$ we mean purely second order perturbations. In the above expressions for completeness we kept all the contributions up to 4th order, however some of them  will be discarded since they do not give contributions to second order perturbation equations (as discussed in the previous section), for example terms like  $\sim (\delta_1)^3$.  
 
In order to obtain the non-linear ADM action, we now apply the prescription of the previous section. In Appendix~\ref{App:Lagrangianexpansion} we report in detail the whole calculation, here we show the final action
\ba \label{actionexpanded}
S_{ADM}^{(4)}&=&\int{}d^4x\sqrt{-g}\l\{\bar{L}+\dot{\mathcal{F}}+3H\mathcal{F}+(L_N-\dot{\mathcal{F}}) \l( \delta_1 N+\f{1}{2}\delta_2N\r)+\l(\dot{\mathcal{F}}+\f{1}{2}L_{NN}\r)\l[(\delta_1 N)^2+\delta_1N\delta_2N+\f{1}{4}(\delta_2 N)^2\r] \r. \nn\\
&&\l.+\f{3}{2}\l(\f{1}{6}L_{NNN}-\dot{\mathcal{F}}\r)(\delta_1N)^2\delta_2N+\l(L_{\mathcal{S}}-3HL_{\tilde{\mathcal{K}}}\r) \l(\delta_1 K_\mu^\nu\delta_1 K_\nu^\mu+\delta_1 K_\mu^\nu\delta_2 K_\nu^\mu+\f{1}{4}\delta_2 K_\mu^\nu\delta_2 K_\nu^\mu\r)\r. \nn\\
&&\l.+\f{1}{2}\mathcal{A}\l[(\delta_1 K)^2+\delta_1K\delta_2K+\f{1}{4}(\delta_2K)^2\r]+\f{1}{4}L_{KKK}(\delta_1K)^2\delta_2K+\mathcal{Q}\l(\delta_1N\delta_1K\delta_2K+\f{1}{2}(\delta_1K)^2\delta_2N\r)\r.\nn\\
&&+\l.\mathcal{B}\l(\delta_1N\delta_1K+\f{1}{2}\delta_1N\delta_2K+\f{1}{2}\delta_2N\delta_1K+\f{1}{4}\delta_2N\delta_2K\r)+\mathcal{I}\l(\f{1}{2}(\delta_1N)^2\delta_2K+\delta_1N\delta_2N\delta_1K\r) \r.\nn\\
&&\l.+\mathcal{C}\l(\delta_1 K \delta_1\mathcal{R}+\f{1}{2}\delta_1K\delta_2\mathcal{R}+\f{1}{2}\delta_2K\delta_1\mathcal{R}+\f{1}{4}\delta_2K\delta_2\mathcal{R}\r)+\mathcal{D}\l[\delta_1 N\delta_1 \mathcal{R}+\f{1}{2}\delta_1N\delta_2\mathcal{R}+\f{1}{2}\delta_2N\delta_1\mathcal{R}+\f{1}{4}\delta_2N\delta_2\mathcal{R}\r]\r.\nn\\
&&\l.+\mathcal{E}\l(\delta_1 \mathcal{R}+\f{1}{2}\delta_2\mathcal{R}\r) +\mathcal{J}\l[\f{1}{2}(\delta_1N)^2\delta_2\mathcal{R}+\delta_1N\delta_2N\delta_1\mathcal{R}\r]+\mathcal{P}\l(\delta_1N\delta_1K^\nu_\mu\delta_2K^\mu_\nu+\f{1}{2}\delta_2N\delta_1K^\nu_\mu\delta_1K^\mu_\nu\r)+\r.\nn\\
&&\l.+\f{1}{2}L_{N\mathcal{U}}\l(\delta_1N\delta_1\mathcal{R}^\mu_\nu\delta_2K^\nu_\mu+\delta_1N\delta_1K^\mu_\nu\delta_2\mathcal{R}^\nu_\mu+\delta_2N\delta_1\mathcal{R}^\mu_\nu\delta_1K^\nu_\mu\r)+L_{K\mathcal{S}}\l(\f{1}{2}\delta_1K^\mu_\nu\delta_1K^\nu_\mu\delta_2 K+\delta_1K^\mu_\nu\delta_2K^\nu_\mu\delta_1K\r)\r.\nn\\
&&\l.+\f{3}{2}L_{\tilde{\mathcal{K}}}\delta_2K_{i}^j\delta_1K^{i}_k\delta_1K_{j}^k+L_{\alpha_1}\l(\delta_1 \alpha_1+\f{1}{2}\delta_2 \alpha_1 \r)+  \f{1}{2}L_{\alpha_1N}\l(\delta_1N\delta_2\alpha_1+\delta_2N\delta_1\alpha_1\r)\r.\nn\\
&&+\l. \f{1}{2}L_{NK\mathcal{R}}  \l( \delta_1N\delta_1K\delta_2\mathcal{R}+\delta_2N\delta_1K\delta_1\mathcal{R} +\delta_1N\delta_2K\delta_1\mathcal{R}\r)
\r\}\,,
\ea
where:~\footnote{Comparing these definitions with the ones in ref.~(\cite{Frusciante:2016xoj}) one can notice some differences. They are simply due to a different way we defined the operators in the original action, mostly due to the introduction of the operator $\tilde{\mathcal{K}}$. Thus, being just a matter of definitions the final results  match each others.}
\begin{align}\label{Coefficientdefinitions}
\mathcal{A}&=L_{KK}-4HL_{SK},\qquad
\mathcal{B}=L_{KN}-2HL_{\mathcal{S}N}+3H^2L_{N\tilde{\mathcal{K}}},\qquad
\mathcal{C}=L_{KR}+\f{1}{2}L_{\mathcal{U}}, \nn\\
\mathcal{D}&= L_{N\mathcal{R}}+\f{1}{2}\dot{L}_{\mathcal{U}}-HL_{N\mathcal{U}}, \qquad
\mathcal{E}=L_{\mathcal{R}}-\f{3}{2}HL_{\mathcal{U}}-\f{1}{2}\dot{L}_{\mathcal{U}}, \qquad
\mathcal{F}=L_K-2HL_{\mathcal{S}}+3H^2L_{\tilde{\mathcal{K}}}, \nn\\
\mathcal{I}&=-HL_{NN\mathcal{S}}+\f{1}{2}L_{NNK}+\f{3H^2}{2}L_{NN\tilde{\mathcal{K}}},\qquad 
\mathcal{J}= -\f{H}{2}L_{NN\mathcal{U}}+\f{1}{2}L_{NN\mathcal{R}}-\f{1}{2}\dot{L}_{\mathcal{U}}, \nn\\
\mathcal{P}&=L_{N\mathcal{S}}-3HL_{N\tilde{\mathcal{\mathcal{K}}}}\,, \qquad
\mathcal{Q}=\f{1}{2}L_{NKK}-2HL_{NK\mathcal{S}}\,.
\end{align}

We have now constructed a general ADM action up to 4th order in perturbations for the relevant operators we need for our purpose. This step is fundamental to construct a general recipe to map any gravity theory in the EFT language. We will show it in  details in the next section.


\subsection{Mapping from a general ADM Lagrangian to the EFT framework}\label{Sec:mapping}


We wish now to proceed and present a self-consistent mapping between the actions~(\ref{EFTfirstaction}) and~(\ref{actionexpanded}) respectively in the EFT framework and the ADM formalism. Once this mapping will have been established it will be fairly easy to embed any single scalar field theory into the EFT framework, as long as it can be rewritten in ADM quantities.

Having expanded in perturbations the ADM action~\eqref{actionexpanded}, now we start to rewrite the EFT action~(\ref{EFTfirstaction}) in terms of the ADM geometric quantities. Once this has been done it will be straightforward to compare the two actions and present the connection between the EFT functions and the derivatives of the Lagrangian present in the ADM action. This procedure follows ref.~\cite{Frusciante:2016xoj} where a similar one has been already employed for the linear EFT action. 

Starting with the Ricci scalar term in action~(\ref{EFTfirstaction}), we get up to 4th-order
\ba
&&\int{}d^4x\sqrt{-g}\f{m_0^2}{2}(1+\Omega)R=\int{}d^4x\sqrt{-g}\l[\f{m_0^2}{2}(1+\Omega)\l(\mathcal{R}+S-K^2\r)+m_0^2\dot{\Omega}\f{K}{N}\r]=\nn\\
&&\hspace{1.8cm}=\int{}d^4x\sqrt{-g}m_0^2\l[\f{1}{2}(1+\Omega)\l(\delta_1\mathcal{R}+\f{1}{2}\delta_2\mathcal{R}\r)+3(1+\Omega)H^2+2\dot{H}(1+\Omega)+2H\dot{\Omega}+\ddot{\Omega}\r.\nn\\
&&\hspace{1.8cm}-\l.\f{1}{2}(1+\Omega)\l((\delta_1K)^2+\delta_1K\delta_2K+\f{1}{4}(\delta_2K)^2-\delta_1K^\nu_\mu\delta_1K^\mu_\nu-\delta_2K^\nu_\mu\delta_1K^\mu_\nu-\f{1}{4}\delta_2K^\nu_\mu\delta_2K^\mu_\nu \r)\r.\nn\\
&&\hspace{1.8cm}+\l.\l(-\dot{\Omega}H+2(1+\Omega)\dot{H}+\ddot{\Omega}\r)\l((\delta_1N)^2-\delta_1N-\f{1}{2}\delta_2N+\f{1}{4}(\delta_2N)^2+\delta_1N\delta_2N-\f{3}{2}(\delta_1N)^2\delta_2N\r)\r.\nn\\
&&\hspace{1.8cm}+\l.\dot{\Omega}\l[\delta_1K\l(-\delta_1N-\f{1}{2}\delta_2N+\delta_1N\delta_2N\r)+\f{1}{2}\delta_2K\l(-\delta_1N+(\delta_1N)^2-\f{1}{2}\delta_2N\r)\r]\r]+\mathcal{O}(5).
\ea
Now, we expand the time-time component of the metric as
\be
g^{00}=-1+\delta_1g^{00}+\f{1}{2}\delta_2 g^{00}=-1+2\delta_1N-3(\delta_1N)^2+\delta_2N-\f{3}{4}(\delta_2N)^2-3\delta_1N\delta_2N+6(\delta_1N)^2\delta_2N+\mathcal{O}(5) \,,
\ee
and obtain the perturbations
\ba
&&\delta g^{00}=2\delta_1N-3(\delta_1N)^2+\delta_2N-\f{3}{4}(\delta_2N)^2-3\delta_1N\delta_2N+6(\delta_1N)^2\delta_2N+\mathcal{O}(5)\,,\nn \\
&&(\delta g^{00})^2=4(\delta_1N)^2+4\delta_1N\delta_2N-18(\delta_1N)^2\delta_2N+(\delta_2N)^2+\mathcal{O}(5)\,.
\ea
Therefore, the new operators  which contribute to second order equations of motion of the scalar perturbations are the following:
\ba
&&(\delta g^{00})^3= 12(\delta_1N)^2\delta_2N\,,\nn\\
&&(\delta K)^3=\f{3}{2}(\delta_1K)^2\delta_2K \,,\nn\\
&&(\delta g^{00})^2\delta K=2(\delta_1N)^2\delta_2K+4\delta_2N\delta_1N\delta_1K\,,\nn\\
&&\delta g^{00}(\delta K)^2=\delta_2N(\delta_1K)^2+2\delta_1N\delta_1K\delta_2K\,,\nn\\
&&(\delta g^{00})^2\delta \mathcal{R}=2(\delta_1N)^2\delta_2\mathcal{R}+4\delta_2N\delta_1N\delta_1\mathcal{R}\,,\nn\\
&&\delta g^{00}\delta K^\mu_\nu\delta K_\mu^\nu=2\delta_1N\delta_1 K^\mu_\nu\delta_2 K_\mu^\nu+\delta_2N\delta_1 K^\mu_\nu\delta_1 K_\mu^\nu\,,\nn\\
&&\delta K\delta K^\mu_\nu\delta K_\mu^\nu=\f{1}{2}\delta_2 K\delta_1 K^\mu_\nu\delta_1 K_\mu^\nu+\delta_1 K\delta_2 K^\mu_\nu\delta_1 K_\mu^\nu\,,\nn\\
&&\delta g^{00}\delta K^\mu_\nu \delta \mathcal{R}^\nu_\mu= \delta_1K^\mu_\nu \delta_1 \mathcal{R}^\nu_\mu \delta_2N+\delta_2K^\mu_\nu \delta_1 \mathcal{R}^\nu_\mu \delta_1N+\delta_1K^\mu_\nu \delta_2 \mathcal{R}^\nu_\mu \delta_1N \,, \nn\\
&&h^{\mu\nu}\partial_\mu g^{00}\partial_\nu g^{00}=4\f{\alpha_1}{N^4}=4\l(\delta_1\alpha_1+\f{1}{2}\delta_2\alpha_1\r)\l(1-4\delta_1N-2\delta_2N \r)\,,\nn\\
&&h^{\mu\nu}\delta g^{00}\partial_\mu g^{00}\partial_\nu g^{00}=4\f{\alpha_1}{N^4}(2\delta_1N+\delta_2N)=4\l(\delta_1\alpha_1\delta_2N+\delta_2\alpha_1\delta_1N\r) \,,
\ea
where we have used for the last two operators the following:
\ba
&&\alpha_1=a^ia_i=h^{ij}\partial_i \ln N \partial_j \ln N=h^{ij}\f{\partial_i N}{N}\f{\partial_j N}{N}\,,\nn\\
&&\partial_{\mu}g^{00}\partial_\nu g^{00}=4\f{\partial_\mu N}{N^3}\f{\partial_\nu N}{N^3}\,.
\ea
The expansion of the other operators is trivial.

Then, the EFT action~(\ref{EFTfirstaction}) in the ADM formalism can be written as follows
\ba\label{EFTADM}
S_{EFT}^{(4)}&=&\int{}d^4x\sqrt{-g}\l\{m_0^2\l[\f{1}{2}(1+\Omega)\l(\delta_1\mathcal{R}+\f{1}{2}\delta_2\mathcal{R}\r)+3(1+\Omega)H^2+2\dot{H}(1+\Omega)+2H\dot{\Omega}+\ddot{\Omega}\r]+\Lambda\r.\nn\\
&-&\l.\l[m_0^2\l(-H\dot{\Omega}+2(1+\Omega)\dot{H}+\ddot{\Omega}\r)+2c\r]\l(\delta_1N+\f{1}{2}\delta_2N\r)\r.\nn\\
&+&\l.\l[m_0^2\l(-H\dot{\Omega}+2(1+\Omega)\dot{H}+\ddot{\Omega}\r)+3c+2M^4_2\r]\l((\delta_1N)^2+\delta_1N\delta_2N+\f{1}{4}(\delta_2N)^2\r)\r.\nn\\
&+&\l.\l[-\f{3}{2}m_0^2\l(-H\dot{\Omega}+2(1+\Omega)\dot{H}+\ddot{\Omega}\r)-6c-9M^4_2+12M^4_3\r]\delta_2N(\delta_1N)^2\r.\nn\\
&+&\l.\l(\f{1}{2}m_0^2(1+\Omega)-\f{\bar{M}^2_3}{2}\r)\l(\delta_1K^\mu_\nu\delta_1K^\nu_\mu+\delta_2K^\mu_\nu\delta_1K^\nu_\mu+\f{1}{4}\delta_2K^\mu_\nu\delta_2K^\nu_\mu\r)\r.\nn\\
&-&\l.\l(\f{1}{2}m_0^2(1+\Omega)+\f{\bar{M}^2_2}{2}\r)\l((\delta_1K)^2+\delta_1K\delta_2K+\f{1}{4}(\delta_2K)^2\r)+\f{3}{2}M_1(\delta_1K)^2\delta_2K\r.\nn\\
&+&\l.\l(4M^2_5-\f{3}{2}\hat{M}^2\r)\l(\f{1}{2}(\delta_1N)^2\delta_2\mathcal{R}+\delta_1N\delta_2N\delta_1\mathcal{R}\r)+\l(m_0^2\dot{\Omega}+\f{3}{2}\bar{M}^3_1+4M^3_1\r)\l(\f{1}{2}(\delta_1N)^2\delta_2K+\delta_1N\delta_2N\delta_1K\r)\r.\nn\\
&+&\l. 2M^2_4\l(\f{1}{2}(\delta_1K)^2\delta_2N+\delta_1N\delta_1K\delta_2K\r)-\l(m_0^2\dot{\Omega}+\bar{M}_1^3\r)\l(\delta_1N\delta_1K+\f{1}{2}\delta_1N\delta_2K+\f{1}{2}\delta_2N\delta_1K+\f{1}{4}\delta_2N\delta_2K\r)\r.\nn\\
&+&\l.\hat{M}^2\l(\delta_1\mathcal{R}\delta_1N+\f{1}{2}\delta_1\mathcal{R}\delta_2N+\f{1}{2}\delta_2\mathcal{R}\delta_1N+\f{1}{4}\delta_2\mathcal{R}\delta_2N\r)+4m^2_2\l(\delta_1\alpha_1+\f{1}{2}\delta_2\alpha_1\r)\r.\nn\\
&+&\l.2M^2_6\l(\delta_1N\delta_1K^\mu_\nu\delta_2K^\nu_\mu+\f{1}{2}\delta_2N\delta_1K^\mu_\nu\delta_1K^\nu_\mu\r)+M_3\l(\delta_1K\delta_1K^\mu_\nu\delta_2K^\nu_\mu+\f{1}{2}\delta_2K\delta_1K^\mu_\nu\delta_1K^\nu_\mu\r)\r.\nn\\
&+&\l.M_4\l(\delta_1N\delta_1\mathcal{R}	\delta_2K+\delta_1N\delta_2\mathcal{R}\delta_1K+\delta_2N\delta_1\mathcal{R}\delta_1K\r)+\f{3}{2}M_2\delta_1K^\mu_\lambda\delta_1K^\lambda_\nu\delta_2K^\nu_\mu \r.\nn\\&+&\l.(4m^2_3-8m^2_2)(\delta_1\alpha_1\delta_2N+\delta_2\alpha_1\delta_1N)
+M_5\l(\delta_1K^\mu_\nu \delta_1 \mathcal{R}^\nu_\mu \delta_2N+\delta_2K^\mu_\nu \delta_1 \mathcal{R}^\nu_\mu \delta_1N+\delta_1K^\mu_\nu \delta_2 \mathcal{R}^\nu_\mu \delta_1N \r)
\r\}\,,
\ea
from which it is easy to identify the following relations with action~(\ref{actionexpanded})
\ba
&&\bar{L}+\dot{\mathcal{F}}+3H\mathcal{F}= m_0^2\l(3(1+\Omega)H^2+2\dot{H}(1+\Omega)+2H\dot{\Omega}+\ddot{\Omega}\r)+\Lambda, \nn\\
&&L_N-\dot{\mathcal{F}}=-\l[m_0^2\l(-H\dot{\Omega}+2(1+\Omega)\dot{H}+\ddot{\Omega}\r)+2c\r], \nn\\
&&\dot{\mathcal{F}}+\f{1}{2}L_{NN}=m_0^2\l(-H\dot{\Omega}+2(1+\Omega)\dot{H}+\ddot{\Omega}\r)+3c+2M^4_2,\nn\\
&&\f{3}{2}\l(\f{1}{6}L_{NNN}-\dot{\mathcal{F}}\r)=-\f{3}{2}m_0^2\l(-H\dot{\Omega}+2(1+\Omega)\dot{H}+\ddot{\Omega}\r)-6c-9M^4_2+12M^4_3,\nn\\
&&L_{\mathcal{S}}-3HL_{\tilde{\mathcal{K}}}=\f{1}{2}m_0^2(1+\Omega)-\f{\bar{M}^2_3}{2},\qquad
\mathcal{A}=-m_0^2(1+\Omega)-\bar{M}^2_2,\qquad \f{1}{4}L_{KKK}=\f{3}{2}M_1,\nn\\
&&\mathcal{Q}= 2M^2_4,\qquad \mathcal{B}=-\l(m_0^2\dot{\Omega}+\bar{M}_1^3\r),\qquad\mathcal{I}= m_0^2\dot{\Omega}+\f{3}{2}\bar{M}^3_1+4M^3_1 \nn\\
&&\mathcal{D}=\hat{M}^2,\qquad \mathcal{E}=\f{1}{2}m_0^2(1+\Omega),\qquad \mathcal{J}=4M^2_5-\f{3}{2}\hat{M}^2,\qquad \mathcal{P}=2M^2_6\nn\\
&&\f{1}{2}L_{NKR}=M_4 ,\qquad L_{K\mathcal{S}}= M_3 ,\qquad L_{\tilde{\mathcal{K}}}=M_2\,, \nn\\
&&\f{1}{2}L_{N\mathcal{U}}=M_5\qquad L_{\alpha_1}=4m^2_2 \qquad \f{L_{\alpha_1N}}{2}=4m^2_3-8m_2^2\,.
\ea
Inverting these relations we obtain the full mapping we set out to derive:
\ba\label{finalmapping}
&&1+\Omega=\f{2\mathcal{E}}{m_0^2}\,,\qquad \Lambda=\bar{L}+\dot{\mathcal{F}}+3H\mathcal{F}-2\l(3\mathcal{E}H^2+2\dot{H}\mathcal{E}+2H\dot{\mathcal{E}}+\ddot{\mathcal{E}}\r)\,,\nn\\
&&c=-\f{1}{2}(L_N-\dot{\mathcal{F}})-\l(-H\dot{\mathcal{E}}+2\mathcal{E}\dot{H}+\ddot{\mathcal{E}}\r)\,, \qquad \bar{M}^2_3=-2L_{\mathcal{S}}+6HL_{\tilde{\mathcal{K}}}+2\mathcal{E}\,,\qquad \bar{M}^2_2=-\mathcal{A}-2\mathcal{E}\,,\nn\\
&&M^4_2=\f{1}{2}\l(L_N+\f{L_{NN}}{2}\r)-\f{c}{2} \,,\qquad \bar{M}^3_1=-2\dot{\mathcal{E}}-\mathcal{B}\,,\qquad \hat{M}^2=\mathcal{D}\,,\qquad m^2_2=\f{L_{\alpha_1}}{4}\,,\nn\\
&&M_1=\f{1}{6}L_{KKK}\,,\qquad M^2_4=\f{1}{2}\mathcal{Q}\,,\qquad M^3_1=\f{1}{4}\l(\mathcal{I}+\dot{\mathcal{E}}+\f{3}{2}\mathcal{B}\r)\,,\qquad M^2_5=\f{1}{4}\mathcal{J}+\f{3}{8}\mathcal{D}\,,\qquad M^2_6=\f{1}{2}\mathcal{P}\,, \nn\\
&&  M_2=L_{\tilde{\mathcal{K}}}\,, \qquad M_3 =L_{K\mathcal{S}}\,, \qquad M_4=\f{1}{2}L_{NKR}\,, \qquad M_5=\f{1}{2}L_{N\mathcal{U}}\,,\nn\\
&& M^4_3=\frac{1}{48} \left(6 \ddot{\mathcal{E}}-6 H \dot{\mathcal{E}}+12 \mathcal{E} \dot{H}-3 \dot{\mathcal{F}}+15 L_N+9 L_{NN}+L_{NNN}\right) \,, \qquad m^2_3=\f{L_{\alpha_1N}}{8}+\f{L_{\alpha_1}}{2}\,.
\ea
The last relations allow to map any action, which has been previously written in ADM form, in the EFT formalism up to the next leading order perturbations. Let us note that the ADM action~(\ref{actionexpanded}) is more general than the EFT one (eq.~(\ref{EFTADM})). For example the ADM action has the term $\mathcal{C}$, which does not appear in the EFT action. That is because in the EFT action such a term corresponds to the EFT function $\bar{m}_5$ (see refs.~\cite{Gleyzes:2013ooa,Frusciante:2016xoj}), which, for the class of models considered in this paper, is zero. We will illustrate the generality of the Lagrangian~(\ref{lagrangianrestricted}) in the Appendix~\ref{Sec:restrictedADMaction}, referring to  all the combinations of perturbative terms which in principle should be present but that have been excluded because the EFT action we are considering is restricted for a class of theories or because some terms have to be excluded in order to have an healthy theory~\cite{Crisostomi:2017aim,BenAchour:2016fzp}.

\subsection{Non-linear beyond Horndeski mapping}\label{Sec:bhmapping}

Now we will apply the procedure described in the previous section to a specific theory, i.e. the beyond Horndeski models~\cite{Gleyzes:2014dya,Gleyzes:2014qga}.

The beyond Horndeski class of models has been presented as an extended version of the Horndeski models allowing for equations of motion which are 3rd order in spatial derivatives. While the additional terms allow for higher order spatial derivatives they have been constructed in such a way that time derivatives do not go beyond the second order. Imposing this restriction it is guaranteed that the theory is free from the Ostrogradski instability~\cite{Ostrogradski} and thus propagates only one additional scalar DoF~\cite{Gleyzes:2014qga,Lin:2014jga,Deffayet:2015qwa}.

For the present purpose, we will consider the Lagrangian as presented in ref.~\cite{Gleyzes:2014dya} as it is directly written in terms of geometrical quantities and it reads
\ba\label{bhlagrangian}
L_{\rm bh}&=&A_2(t,N)+A_3(t,N)K+A_4(t,N)(K^2-K_{ij}K^{ij})+B_4(t,N)\mathcal{R}\nn\\
&+&A_5(t,N)\l(K^3-3KK_{ij}K^{ij}+2K_{ij}K^{ik}K^{j}_k\r)+B_5(t,N)K^{ij}\l(\mathcal{R}_{ij}-h_{ij}\f{\mathcal{R}}{2}\r)\,,
\ea
where $A_{i}, B_{i}$ are general functions of $t$ and $N$. The above Lagrangian can be rewritten in terms of the scalar field, $\phi$, as shown in ref.~\cite{Gleyzes:2014dya}. The way the Lagrangian~(\ref{bhlagrangian}) is written greatly simplifies the steps to write it in term of the operators introduced in Sec.~\ref{Sec:restrictedADMaction}, indeed we have:
\ba
L_{\rm bH}&=&A_2(t,N)+A_3(t,N)K+A_4(t,N)(K^2-\mathcal{S})+B_4(t,N)\mathcal{R}\nn\\
&+&A_5(t,N)\l(K^3-3K\mathcal{S}+2\tilde{\mathcal{K}}\r)+B_5(t,N)\l(\mathcal{U}-\f{K\mathcal{R}}{2}\r)\,,
\ea
with which it is easy to apply the prescription in eqs.~(\ref{finalmapping}) and obtain the new operators
\ba
&&M_2=2A_5 \,,\quad M_3=-3A_5 \,,\quad M_4=-\f{1}{2}B_{5N}\,, \quad M_1=A_5,\quad M^2_4=\f{1}{2}\l(A_{4N}-3HA_{5N}\r) \,, \quad M_5=\f{1}{2}B_{5N}\,,\nn\\
&& M^2_6=\f{1}{2}(3HA_{5N}-A_{4N})\,, \quad M^2_5=\f{1}{8}\l(\f{1}{2}HB_{5NN}+B_{4NN}+\f{1}{2}\dot{B}_5+3B_{4N}+\f{3}{2}HB_{5N}\r),\nn\\
&&M^3_1=\f{1}{4}\l(-2HA_{4NN}+\f{1}{2}A_{3NN}+3H^2A_{5NN}+\dot{B}_4-\f{1}{2}\ddot{B}_5+\f{3}{2}A_{3N}-6HA_{4N}+9H^2A_{5N}\r),\nn\\
&&M^4_3= \f{1}{48}\l[A_{2NNN}-3HA_{3NNN}+6H^2A_{4NNN}-6H^3 A_{5NNN}-3(\dot{A}_3-4\dot{H}A_4-4H\dot{A}_4+6H^2 \dot{A}_5+12 H \dot{H}A_5)\r.\nn\\
&&\l.+15\l( A_{2N}-3HA_{3N}+6H^2A_{4N}-6H^3 A_{5N}\r)+9\l( A_{2NN}-3HA_{3NN}+6H^2A_{4NN}-6H^3 A_{5NN}\r) \r.\nn\\
&&\l.+6\l(\ddot{B}_4-\f{1}{2}B_5^{(3)}\r)-6H\l(\dot{B}_4-\f{1}{2}\ddot{B}_5\r)+12\dot{H}\l(B_4-\f{1}{2}\dot{B}_5\r)\r]\,,
\ea
and  $m^2_3=0$. The mapping of the linear operators can be found in ref.~\cite{Frusciante:2016xoj}. As one can notice from the mapping in eq.~(\ref{bhlagrangian})
many EFT functions are not independent, rather we have the following relations
\be
M^2_6=-M^2_4\,, \qquad 6M_1=3M_2=-2M_3\,, \qquad M_5=-M_4\,.
\ee 
As a consequence, one needs only 6 independent new EFT functions to describe beyond Horndeski at second order in perturbations.


\subsection{Non linear low-energy Ho\v rava gravity mapping}\label{Sec:horavamapping}


A second example we will use in this section is the low-energy Ho\v rava gravity~\cite{Horava:2008ih,Horava:2009uw} and we will provide the mapping for the new EFT functions.

In this theory the action is modified by adding higher order spatial derivatives but without adding higher order time derivatives to avoid Ostrogradski instabilities~\cite{Ostrogradski}.  From a practical point of view the theory has been constructed by considering  space and time on different footing thus leading to the breaking of  the full diffeomorphism invariance and to Lorentz violations at all scales. In this way the theory propagates one extra scalar DoF. Moreover, the theory is  renormalizable, thanks to power-counting arguments~\cite{Visser:2009fg,Visser:2009ys}, as expected for a candidate for quantum gravity. 

Here, we will consider the low-energy Ho\v rava gravity action, which is constructed with all the operators satisfying the above requirements and with second order spatial derivatives. Then, the  action  reads~\cite{Blas:2009qj}:
\begin{eqnarray}\label{actionhorava}
\mathcal{S}_{H}=\f{1}{16\pi G_H}\int{}d^4x\sqrt{-g}\left[K_{ij}K^{ij}-\lambda K^2 -2 \xi\bar{\Lambda}+\xi \mathcal{R}+\eta a_i a^i \right],
\end{eqnarray}
where  the coefficients $\lambda$, $\eta$, $\xi$ are running coupling constants,  $\bar{\Lambda}$ is the ''bare'' cosmological constant and $G_H$ is the coupling constant~\cite{Blas:2009qj,Frusciante:2015maa}:
\be
\f{1}{16\pi G_H}=\f{m_0^2}{(2\xi-\eta)}\,.
\ee
Again, also in this case it will be very easy to translate the above action in terms of the operators presented in Sec.~\ref{Sec:mapping}, because the action is already written in ADM formalism. Then, we get
\begin{eqnarray}
\mathcal{S}_{H}=\f{m_0^2}{(2\xi-\eta)}\int{}d^4x\sqrt{-g}\left[\mathcal{S}-\lambda K^2 -2 \xi\bar{\Lambda}+\xi \mathcal{R}+\eta \alpha_1\right].
\end{eqnarray}
Now, using the prescription in eqs.~(\ref{finalmapping}) it is easy to show that the new EFT functions are:
\be
M^4_3=\f{m_0^2\dot{H}}{8(2\xi-\eta)}\l[1+2\xi-3\lambda\r]\,,\qquad m^2_3=\f{m_0^2\eta}{2(2\xi-\eta)} \,,
\ee
being the others zero. The mapping of the linear part of the action in terms of the EFT functions can be found in ref. ~\cite{Frusciante:2015maa,Frusciante:2016xoj}. 

Finally, in the case of low-energy Ho\v rava gravity one needs to account for two extra new EFT functions for the next to leading order in the expansion.


\section{Quartic action and stability}\label{Sec:4actionstability}


In this section we will  consider for the first time the 4th order action in terms of non-linear perturbation from which it will be possible to obtain the non linear perturbed equation for the curvature perturbation at the next to leading order. The derivations has been done by using the \textit{Mathematica} packages \textit{xAct}\cite{xAct} and \textit{xPand}\cite{xPand}.  Once this has been done we will conclude by  commenting on the stability of the second order perturbation. 

The second order action has been largely considered in refs.~\cite{Gubitosi:2012hu,Bloomfield:2012ff,Bloomfield:2013efa,Gleyzes:2013ooa,Gleyzes:2014rba,Frusciante:2015maa,Frusciante:2016xoj,DeFelice:2016ucp}, where  the linear perturbed equations and the stability conditions have been derived and  the phenomenology associated to the extra d.o.f. has been investigated. We will mention it in the following just for completeness. In what follows we will restrict our analysis only to the class of theories belonging to beyond Horndeski, the generalization is quite straightforward.

Let us start by  considering the perturbations of the ADM metric components (eq.~\ref{ADMmetric}):
\ba \label{perturbedmetric}
&&N= 1+\delta_1 N+\f{1}{2}\delta_2 N+ \mathcal{O}(3)\,,\nn\\
&&N_i =\psi_1 +\f{1}{2}\psi_2+ \mathcal{O}(3)\,,\nn\\
&&\gamma_{ij}=a^2e^{2\zeta}=\delta_{ij}a^2\l(1+2\zeta_1+2\zeta_1^2+\zeta_2+\f{1}{2}\zeta_2^2+2\zeta_1\zeta_2+2\zeta_1^2\zeta_2\r)+ \mathcal{O}(5)\,.
\ea
We are considering only the scalar part of the metric while neglecting vector and tensor perturbations. The reason is the following~\cite{Matarrese:1997ay,Bartolo:2013ws}: at linear order scalar and tensor perturbations completely decouple and vector perturbations decay. At the next to leading order tensor perturbations might appear in the second order scalar equation but their contribution is negligible with respect to the linear scalar perturbation contribution.  The opposite does not hold in general, i.e. linear order scalar perturbations cannot be neglected in the second order equations for vector and tensor fields. Since in this section we will focus only on the 4th order action for scalar perturbations, for the reasons mentioned above, we will neglect  the linear vector and tensor perturbations contributions.  

The second order action for the linear scalar perturbations, after integrating out the non dynamical fields ($\delta N_1, \psi_1$) and considering the Fourier Transform of the spatial part \footnote{We have considered the following Fourier Transform for the perturbation functions 
\begin{equation}
\phi(\vec{x},t)=\f{1}{(2\pi)^3}\int dk^3 \phi(\vec{k},t) e^{i \vec{k}\cdot \vec{x}}\,,
\end{equation}
however in the text we dropped the vector form on $\vec{k} \rightarrow k$. }, reads~\cite{Frusciante:2016xoj}
\begin{eqnarray}\label{actionsecondorder}
S^{(2)}&=&\int \f{dk^3}{(2\pi)^3} a^3\l\{\l[\f{3}{2}\mathcal{W}_5+\f{\mathcal{W}_1\mathcal{W}_5^2}{\mathcal{W}_4^2}\r]\dot{\zeta}_1(k)\dot{\zeta}_1(-k)-\f{k^2}{a^2}\l[\f{H}{2}\f{\mathcal{W}_5\mathcal{W}_6}{\mathcal{W}_4}+\f{1}{2}\f{d}{dt}\l(\f{\mathcal{W}_5\mathcal{W}_6}{\mathcal{W}_4}\r)\r]\zeta_1(k)\zeta_1(-k))\r\}\,,
\end{eqnarray}
where we have defined
\ba
&&\mathcal{W}_0=-m_0^2(1+\Omega)\,,\nn \\
&&\mathcal{W}_1= c+2M^4_2-3m_0^2H^2(1+\Omega)-3m_0^2H\dot{\Omega}-3H^2\bar{M}^2_2-3H\bar{M}^3_1\,, \nn \\
&&\mathcal{W}_4= -2m_0^2H(1+\Omega)-m_0^2\dot{\Omega}-\bar{M}^3_1-2H\bar{M}^2_2 \,,\nn\\ 
&&\mathcal{W}_5=  2m_0^2(1+\Omega) +2\bar{M}^2_2\,,\nn\\
&&\mathcal{W}_6= -4\l(\f{1}{2}m_0^2(1+\Omega)+\hat{M}^2\r)\,.
\ea
Let us now focus on the 4th order action and write it as follows
\begin{eqnarray}
S^{(4)}=S^{(4)}_{22}+S^{(4)}_{21}\,,
\end{eqnarray}
where $S^{(4)}_{22}$ is the 4th order action made by purely second order perturbations and $S^{(4)}_{21}$ contains 4th order terms resulting from the coupling between linear perturbations and purely second order ones. Once all the operators in action (\ref{EFTnonlinear}) have been written in terms of perturbations given by the perturbed metric (\ref{perturbedmetric}), $S^{(4)}_{22}$  and $S^{(4)}_{12}$  read
\begin{eqnarray}
S^{(4)}_{22}&=&\int \f{dk^3}{(2\pi)^3} a^3\l\{-\mathcal{W}_0\f{k^2}{a^2}\zeta_2(k)\zeta_2(-k)-3\mathcal{W}_4 \delta N_2(k)\dot{\zeta}_2(-k)-\f{3}{2}\mathcal{W}_5\dot{\zeta}_2(k)\dot{\zeta}_2(-k)\r.\nn\\
&-&\l.\l[\mathcal{W}_4 N_2(k)+\mathcal{W}_5\dot{\zeta}_2(k)\r]\f{k^2}{a^2}\psi_2(-k)+\mathcal{W}_1\delta N_2(k)\delta N_2(-k) -\mathcal{W}_6\delta N_2(k)\zeta_2(-k)\f{k^2}{a^2}\r\}\,,
\end{eqnarray}
and
\begin{eqnarray}
S^{(4)}_{21}&=&\int \f{dk^3dk_1^3}{(2\pi)^6} a^3\l\{X_1 \delta N_2(-k)\delta N_1(k_1)\delta N_1(k-k_1)+\l(3\mathcal{W}_1+2X_3\f{k_{1}^2}{a^2}\r) \delta N_2(-k)\zeta_1(k_1)\delta N_1(k-k_1)\r.\nn\\
&&\l. +\mathcal{W}_6\f{(k_1^2+k\cdot k_1)}{4a^2}\delta N_2(-k)\zeta_1(k_1)\zeta_1(k-k_1)+\l(\f{3\mathcal{W}_1}{2}+X_3\f{k^2}{a^2}\r)\zeta_2(-k)\delta N_1(k_1)\delta N_1(k-k_1)\r.\nn\\
&&\l.+\mathcal{W}_6\l(-\f{k^2}{2a^2}-\f{k_1^2}{2a^2}+\f{k\cdot k_1}{2a^2}\r) \zeta_2(-k)\zeta_1(k_1)\delta N_1(k-k_1)-\mathcal{W}_0\f{(k^2+k_1^2-k\cdot k_1)}{2a^2}\zeta_2(-k)\zeta_1(k_1)\zeta_1(k-k_1) \r.\nn\\
&&\l. +X_2\f{k_1^2}{2a^2}\delta N_2(-k)\psi_1(k_1)\delta N_1(k-k_1)+\l( M_4\f{4 (k\cdot k_1)^2 -4k^2 k_1^2}{a^4} +\mathcal{W}_4\f{k_1^2-k\cdot k_1}{2a^4}\r)\zeta_2(-k)\psi_1(k_1)\delta N_1(k-k_1)\r.\nn\\
&&\l.+\l(\f{4\l(\l(k-k_1\r)\cdot k_1\r)^2-4k_1^2(k-k_1)^2)}{a^4}-\f{k_1^2\mathcal{W}_4+(k-k_1)\cdot k_1 \mathcal{W}_4}{2a^2}\r)\delta N_2(-k)\psi_1(k_1)\zeta_1(k-k_1)\r.\nn\\
&&\l.+\l(\delta N_1(k_1)\l(-\f{8k^2M_4}{a^2}-\f{9}{2}\mathcal{W}_4\r)+\psi_1(k_1)\l(\f{\mathcal{W}_5(k\cdot k_1-k_1^2)}{2a^2}\r)\r)\zeta_2(-k)\dot{\zeta}_1(k-k_1)\r.\nn\\
&&\l.+\l( 6X_2\delta N_1(k_1)-\zeta(k_1)\l( \f{8M_4 k_1^2}{a^2}+\f{9\mathcal{W}_4}{2} \r)-\f{k^2 X_8}{a^2}\psi_1(k_1)\r)\delta N_2(-k)\dot{\zeta}(k-k_1)\r.\nn\\
&&\l.+\l(\f{X_{10}(k_1\cdot\l(k-k_1\r))^2+X_4k_1^2(k-k_1)^2}{2a^4} \r) \delta N_2(-k)\psi_1(k_1)\psi_1(k-k_1)+X_2\f{k^2}{a^2}\psi_2(-k)\delta N_1(k_1)\delta N_1(k-k_1)\r.\nn\\
&&\l.+\mathcal{W}_5\l(\f{k_1^2k^2-16k_1^2 k\cdot k_1+3(k\cdot k_1)^2}{8a^4} \r)\zeta_2(-k)\psi(k_1)\psi_1(k-k_1)+\l( 2X_2 \delta N_1(k_1)\delta N_1(k-k_1)\r)\dot{\zeta}_2(-k)\r.\nn\\
&&\l. +\l[-\l(\f{8k_1^2 M_4+\f{9W_4}{2}}{a^2} \r)\zeta_1(k_1)\delta N_1(k-k_1) -3X_8 \delta N_1(k_1)\dot{\zeta}_1(k-k_1)-\f{9\mathcal{W}_5}{2}\zeta_1(k_1)\dot{\zeta}_1(k-k_1)\r.\r.\nn\\
&&\l.\l.-X_8\f{k_1^2}{a^2}\psi_1(k_1)\delta N_1(k-k_1)-\mathcal{W}_5\l( \f{k_1^2+(k-k_1)\cdot k_1}{2a^2} \r)\psi_1(k_1)\zeta_1(k-k_1)-\f{54k_1^2M_1}{a^2}\psi_1(k_1)\dot{\zeta}_1(k-k_1)\r.\r.\nn\\
&&\l.\l.+\l( \f{3k_1^2(k-k_1)^2M_1 -12((k-k_1)\cdot k_1)^2M_1}{2a^4} \r)\psi_1(k_1)\psi_1(k-k_1)-36M_1 \dot{\zeta}_1(k_1)\dot{\zeta}_1(k-k_1) \r]\dot{\zeta}_2(-k)
\r.\nn\\
&&\l.+\mathcal{W}_5\l(\f{k^2k_1^2-2(k\cdot k_1) k\cdot (k-k_1)-(k\cdot k_1)^2+2(k\cdot k_1) (k-k_1)\cdot k_1 }{2a^4}   \r) \psi_2(-k)\psi_1(k_1)\zeta_1(k-k_1)\r.\nn\\
&&\l. +\l( -\f{X_8 k^2}{a^2}\delta N_1(k_1)+\mathcal{W}_5\l( \f{k\cdot k_1-k^2}{2a^2} \r)\zeta_1(k_1)+3M_1\l(\f{k^2k_1^2+(k\cdot k_1)^2}{a^4}\r) \psi_1(k_1)  \r)\psi_2(-k)\dot{\zeta}_1(k-k_1)\r.\nn\\
&&\l.+\l( 4M_4\f{(k\cdot k_1)-k^2k_1^2}{a^4}  -\mathcal{W}_4\f{k^2+k\cdot k_1}{2a^2} \r)\psi_2(-k)\zeta_1(k_1)\delta N_1(k-k_1)
+\f{1}{2a^6}\l[ 3k^2k_1^2(k-k_1)^2\r.\r.\nn\\
&&\l.\l.-6k_1^2 (k\cdot (k-k_1))^2-3 k^2(k_1\cdot (k-k_1))+6(k\cdot k_1)\big(k\cdot(k-k_1)\big)\big(k_1\cdot(k-k_1)\big) \r]\psi_2(-k)\psi(k_1)\psi(k-k_1)
\r.\nn\\
&&\l.-\l( X_8 \delta N_2(-k)\f{3}{2}+\f{32k^2M_1}{3a^2}\psi_2(-k) \r)\dot{\zeta}_1(k_1)\dot{\zeta}_1(k-k_1)\r.\nn\\
&&\l.+\l(\f{X_{10} (k\cdot k_1)^2+X_4 k^2 k_1^2}{a^4}\r)\psi_2(-k)\psi_1(k_1)\delta N_1(k-k_1)\r\}\,,
\end{eqnarray}
with the following definitions
\begin{eqnarray}
&&X_1=\f{3}{4}\Big[ -2c-16H^3M_3-8(M_2^4-2M_3^4)+3H(8M_1^3+3\bar{M}_1^3+2m_0\dot{\Omega})+3H^2(2\bar{M}_2^2+8M_4^2+m_0^2(1+\Omega))  \Big]\,,\nn\\
&&X_2=-\f{3}{4}\Big[ -16H^2M_3+8M_1^3+3\bar{M}_1^3+2H[2\bar{M}^2_2+2(4M_4^2+m_0^2(1+\Omega))]  \Big]\,,\nn\\
&&X_3=8M_5^2-\hat{M}^2\,,\nn\\
&&X_4=\f{1}{2}\Big[ m_0^2(1+\Omega)+\bar{M}^2_2+4M^4_2+6HM_1  \Big]\,,\nn\\
&&X_8=m_0^2(1+\Omega)+\bar{M}^2_2+4M_4^2-16HM_3\,,\nn\\
&&X_{10}=-\f{\mathcal{W}_5}{4} -2M_4^2 + H M_3\,.
\end{eqnarray}

Now we  proceed to remove the non dynamical fields $\{\delta N_2, \delta N_1, \psi_1,\psi_2\}$ from the action $S^{(4)}$, resulting in an action dependent only on the curvature perturbation $\{\zeta_2,\zeta_1\}$ and their derivatives. First, in order to eliminate the $\{\delta N_1,\psi_1\}$ fields we will use the constraint equations from the second order action (see ref. \cite{Frusciante:2016xoj} for details), which have been used to obtain the final form for action (\ref{actionsecondorder}). They will introduce in action $S^{(4)}$ terms proportional to $\{\dot{\zeta}_1, \zeta_1\}$. At this point it is possible to vary the 4th order action with respect to $\delta N_2$ and $\psi_2$ and use the resulting constraint equations to eliminate these fields from the action. After some manipulation we end up with the following two parts for the action $S^{(4)}$:  
\begin{eqnarray}\label{action22final}
S^{(4)}_{22}&=&\int \f{dk^3}{(2\pi)^3} a^3\l\{\l(\f{3}{2}\mathcal{W}_5+\f{\mathcal{W}_1\mathcal{W}_5^2}{\mathcal{W}_4^2}\r)\dot{\zeta}_2(k)\dot{\zeta}_2(-k)-\f{k^2}{a^2}\l[\f{H}{2}\f{\mathcal{W}_5\mathcal{W}_6}{\mathcal{W}_4}+\f{1}{2}\f{d}{dt}\l(\f{\mathcal{W}_5\mathcal{W}_6}{\mathcal{W}_4}\r)\r]\zeta_2(k)\zeta_2(-k))\r\}\,,
\end{eqnarray}
and
\begin{eqnarray} \label{action12final}
S^{(4)}_{21}=&&\int \f{d^3k d^3k_1}{(2\pi)^6}a^3\l[ \zeta_2(-k)\l\{ K_1(k,k_1)\zeta_1(k_1)\zeta_1(k-k_1)+K_2(k,k_1)\dot{\zeta}_1(k_1)\zeta_1(k-k_1)+K_3(k,k_1)\zeta_1(k_1)\dot{\zeta_1}(k-k_1)\r.\r.\nn\\
&&\l.\l.+K_4(k,k_1)\dot{\zeta}_1(k_1)\dot{\zeta}_1(k-k_1) \r\}+\dot{\zeta}_2(-k)\l\{  K_{1d}(k,k_1)\zeta_1(k_1)\zeta_1(k-k_1)+K_{2d}(k,k_1)\dot{\zeta}_1(k_1)\zeta_1(k-k_1)\r.\r.\nn\\
&&\l.\l.+K_{3d}(k,k_1)\zeta_1(k_1)\dot{\zeta_1}(k-k_1)+K_{4d}(k,k_1)\dot{\zeta}_1(k_1)\dot{\zeta}_1(k-k_1)\r\} \r]\,.
\end{eqnarray}
Due to their complicated nature we will present the Kernels $K_i, K_{i d}$ in Appendix \ref{App:Kernels}. The variation of the action $S^{(4)}$ with respect to $\zeta_2$ thus gives the non linear dynamical evolution equation for the curvature perturbation at the second perturbative order.  Let us stress that the action presented in Eqs. (\ref{action22final})-(\ref{action12final})  is crucial for the study of the mildly non linear regime and for the investigation of the impact that modifications of gravity might have on the observables.

From the structure of actions (\ref{action22final})-(\ref{action12final}), it is possible to deduce that the linear order perturbations appearing in action (\ref{action12final}) can be interpreted as  source terms  modifying the evolution of the second order curvature perturbation. Moreover these terms introduce non linearities in the equations thus are responsible for any effect due to screening mechanisms.    

The stability conditions can be read off from the action (\ref{action22final}) which as expected assumed the same form of the second order action (\ref{actionsecondorder}). Thus the stability conditions for the avoidance of ghosts in the scalar sector and the condition on the positivity of the speed of propagation for the scalar mode at next leading order do not change. We write  them here for completeness \cite{Frusciante:2016xoj}:
\begin{eqnarray}
&&\f{3}{2}\mathcal{W}_5+\f{\mathcal{W}_1\mathcal{W}_5^2}{\mathcal{W}_4^2}>0 \,,\nn\\
 &&c_s^2=\f{\f{H}{2}\f{\mathcal{W}_5\mathcal{W}_6}{\mathcal{W}_4}+\f{1}{2}\f{d}{dt}\l(\f{\mathcal{W}_5\mathcal{W}_6}{\mathcal{W}_4}\r)}{\f{3}{2}\mathcal{W}_5+\f{\mathcal{W}_1\mathcal{W}_5^2}{\mathcal{W}_4^2}}>0\,.
\end{eqnarray}

Let us note that despite the results in this section have been derived for the beyond Horndeski class of models they can be easily extended to Lorentz violating theories, which are the only ones excluded.  In particular, we do not expect that any extension of the above treatment to more general theories will change the stability requirement  with respect to the one obtained for linear theory in ref. \cite{Frusciante:2016xoj}.


\section{Conclusions}\label{Sec:conclusion}


Identifying the correct underlying theory of gravity comes more and more within our reach as high precision observational cosmology develops. However, connecting theory with observations or N-body simulations remains a difficult task. Usually, Boltzmann codes or N-body codes rely on specific gravity theories, rendering the testing of a large number of models heavily resource consuming. Recently, tools to study linear cosmological perturbations in a quite general fashion have been proposed \cite{Hu:2013twa,Raveri:2014cka,Zumalacarregui:2016pph,Huang:2012mt}, including one based on the EFT of DE/MG \cite{eftcamb}. The EFT formalism presents a unifying and model independent framework  to study linear perturbations of a large class of single scalar-tensor theories. 

In this paper we presented the extension of the EFT of DE/MG to the next to leading order in perturbations while preserving the model independent aspect which is typical of the EFT formalism. This extension will allow for the study of corrections to the power spectrum coming from non-linearities as well as high order correlators such as the bispectrum, opening up the possibility to study the mildly non-linear scales ($k\gtrsim 0.1h/Mpc$) of a wide class of MG and DE models. These, intermediate scales, are of particular interest as a substantial part of current and upcoming cosmological datasets come from there, hence, being able to extract information at this regime will improve our ability to test theories of gravity.

In order to present the extended action~\eqref{EFTnonlinear}, we proceeded to identify the necessary operators which have to be included in the EFT framework when considering higher order perturbations. This was not a trivial task as to go one order higher in the equations of motion requires one to go up to the 4th-order in perturbations in the action. Many relevant operators can in principle be constructed and added to the Lagrangian, therefore we decided to focus on those which are necessary to expand to the next to leading order in perturbations the class of theories to which beyond Horndeski and low-energy Ho\v rava belong. Let us notice that in doing this choice we are covering most of the theories which are of cosmological interest.  We left for future work the inclusion of a larger class of theories to the extended EFT action, such as high-energy Ho\v rava gravity~\cite{Blas:2009qj} and DHOST~\cite{Langlois:2017mxy}.  In this respect, we have identified 11 new operators each with its own EFT function. 

Having extended the EFT framework with the necessary operators we proceeded in constructing a general and self consistent recipe which maps any given theory (which falls in the beyond Horneski or low-energy Ho\v rava gravity classes) into the EFT formalism. In order to obtain such recipe, we started by writing a general Lagrangian in terms of ADM quantities. This led to the identification of all the operators contributing to the next to linear order in perturbations and with up to six spatial derivatives. Subsequently, we restricted our analysis only to those contributing to beyond Horndeski and low-energy Ho\v rava gravity classes and we have expanded the Lagrangian for the relevant operators up to 3rd order in the total perturbations. To complete the mapping we had to use a similar procedure in order to rewrite the EFT Lagrangian in terms of the geometrical ADM quantities. Finally, we proceeded to relate the EFT Lagrangian to the ADM one. As a result, each EFT function is written in terms of the general Lagrangian thus simplifying the mapping of a chosen theory in the EFT framework. This exemplifies the EFT as a unifying framework, besides its model independent qualities. We concluded with the application of the mapping on the two main classes of theories which inspired the EFT action, the beyond Horndeski theory and the low-energy Ho\v rava one. We found that in the case of the beyond Horndeski models only 6 new functions out of the original 11 needs to be considered for second order perturbations equations, while in the case of low- energy Ho\v rava, they drastically reduce to two.

Finally, we have derived the non-linear action for the curvature perturbation which is a novel result in the field for
the mildly non linear regime which will prove useful when investigating the behavior of second order perturbations and their impact on observables. We found that the purely second order part of the action resembles the one at linear order, plus a piece which account for the interaction between linear and non linear terms. In particular this part of the action is responsible of the non linearities in the equation for the second order curvature perturbation because it will act as a source term.  The structure of the action  confirms that the stability conditions for the scalar mode (i.e. avoidance of ghost instability and positive speed of propagation) derived at the linear level guarantee the stability of higher orders as well. Thus, imposing a stable linear theory will ensure the stability at the next perturbative order as well as expected.

In future works as a first application of this newly developed non-linear formalism, we will proceed to calculate the model independent matter power spectrum going one order beyond the linear one.

\begin{acknowledgments}
We are grateful to Alessandra Silvestri for useful discussions and for comments on the manuscript.
	 The research of NF is   supported by Funda\c{c}$\tilde{\textit{a}}$o para a
  Ci$\hat{\textit{e}}$ncia e a Tecnologia (FCT) through national funds
  (UID/FIS/04434/2013) and by FEDER through COMPETE2020
  (POCI-01-0145-FEDER-007672).  GP acknowledges support from the D-ITP
  consortium, a program of the Netherlands Organisation for Scientific
  Research (NWO) that is funded by the Dutch Ministry of Education,
  Culture and Science (OCW). NF and GP acknowledge the COST Action
  (CANTATA/CA15117), supported by COST (European Cooperation in
  Science and Technology).
\end{acknowledgments}

\appendix

\section{Expanding the ADM Lagrangian}\label{App:Lagrangianexpansion}

In this appendix we present the non-linear ADM Lagrangian  we have used to derive the action in Sec~\ref{Sec:restrictedADMaction}. 
 
According to the expansion~(\ref{Lagrangianexpansion}) up to 3rd order for the operators in the Lagrangian~(\ref{lagrangianrestricted}) we can write
\ba
L&=&\bar{L}+L_N\delta N+L_K\delta K+ L_\mathcal{S}\delta \mathcal{S}+L_\mathcal{U}\delta \mathcal{U}+L_{\tilde{\mathcal{K}}} \delta \tilde{\mathcal{K}}+L_\mathcal{R}\delta \mathcal{R}+L_{\alpha_1} \delta \alpha_1+\f{1}{2}\l[L_{NN}\delta N^2+L_{KK}\delta K^2\r.\nn\\&+&\l.2\l(L_{NK}\delta N\delta K+L_{N\mathcal{S}}\delta N\delta \mathcal{S}+L_{N\mathcal{U}}\delta N\delta \mathcal{U}+L_{N\mathcal{R}}\delta N \delta \mathcal{R}+L_{N\alpha_1}\delta N\delta \alpha_1+L_{K\mathcal{S}}\delta K\delta \mathcal{S}\r.\r.\nn\\
&+&\l.\l.L_{K\mathcal{R}}\delta K \delta \mathcal{R}\r)\r]+\f{1}{6}\l( L_{NNN}\delta N^3+L_{KKK}\delta K^3 \r)+L_{K\mathcal{R}N}\delta N\delta K\delta \mathcal{R}+L_{KN\mathcal{S}}\delta N \delta K \delta \mathcal{S} \nn\\
&+&\f{1}{2}\l(L_{NNK}\delta N^2\delta K+L_{NN\tilde{\mathcal{K}}}\delta N^2\delta \tilde{\mathcal{K}}+L_{NN\mathcal{S}}\delta N^2\delta \mathcal{S}+L_{NN\mathcal{U}}\delta N^2\delta \mathcal{U}+L_{NN\mathcal{R}}\delta N^2 \delta \mathcal{R}+L_{KKN}\delta K^2\delta N\r)\,.
\ea
Let us note that we have excluded from the above expansion all the terms which are not included in the EFT action~(\ref{EFTnonlinear}), such as $L_{\mathcal{S}\mathcal{S}}\delta\mathcal{S}^2, L_{\mathcal{U}\mathcal{U}}\delta \mathcal{U}^2,L_{KK\mathcal{S}}\delta K^2\delta \mathcal{S}, L_{K\alpha_1} \delta K \delta\alpha_1,L_{\mathcal{S}\mathcal{U}}\delta \mathcal{S}\delta \mathcal{U},L_{\mathcal{S}\mathcal{R}}\delta \mathcal{S} \delta \mathcal{R},L_{\mathcal{R}\mathcal{U}}\delta \mathcal{R}\delta \mathcal{U}, L_{\tilde{\mathcal{K}}\tilde{\mathcal{K}}} \delta \tilde{\mathcal{K}}^2,L_{\mathcal{R}\mathcal{R}}\delta \mathcal{R}^2,L_{K\mathcal{U}}\delta K\delta \mathcal{U}$. Indeed, in order to include some of these operators the EFT action should be extended as well by considering additional operators, such as  $\lambda_1(t)\delta \mathcal{R}^2$ which will account for $\delta \mathcal{R}^2$, $\delta \mathcal{U}^2$ and $\delta \mathcal{R} \delta \mathcal{U}$.  In this respect the Lagrangian described by the operators $\{N, K,\mathcal{S},\mathcal{R},\mathcal{U},\tilde{\mathcal{K}},\alpha_1\}$ is more general than the EFT action. However, let us note that not all the combinations can be considered in general since some of them might imply an unhealthy theory~\cite{Crisostomi:2017aim,BenAchour:2016fzp}. 

In order to obtain action~(\ref{actionexpanded}) we have adopted the following expansions: 
\ba
&&L_N\delta N=L_N(\delta_1N+\f{1}{2}\delta_2N)\,,\nn \\
&&\f{1}{2}L_{NN}(\delta N)^2=\f{1}{2}L_{NN}\l[(\delta_1N)^2+\delta_1N\delta_2N+\f{1}{4}(\delta_2N)^2\r] +\mathcal{O}(5)\,,\nn\\
&&\f{1}{6}L_{NNN}(\delta N)^3=\f{1}{6}L_{NNN}\l[ (\delta_1N)^3+\f{3}{2}(\delta_1N)^2\delta_2N \r] \,,\nn+\mathcal{O}(5) \\
&& L_{N\mathcal{S}}\delta N\delta \mathcal{S}=-2HL_{N\mathcal{S}}\l(\delta_1N\delta_1K+\f{1}{2}\delta_1N\delta_2K+\f{1}{2}\delta_2N\delta_1K+\f{1}{4}\delta_2N\delta_2K\r)\nn\\
&&\,\,\,+L_{N\mathcal{S}}\l[\delta_1N(\delta_1 K^\nu_\mu \delta_1 K^\mu_\nu+ \delta_1 K^\nu_\mu \delta_2 K^\mu_\nu)+\f{1}{2}\delta_2N\delta_1 K^\nu_\mu \delta_1 K^\mu_\nu\r]+\mathcal{O}(5)\,,\nn\\
&&L_{NK}\delta N\delta K=L_{NK}\l(\delta_1N\delta_1K+\f{1}{2}\delta_1N\delta_2K+\f{1}{2}\delta_2N\delta_1K+\f{1}{4}\delta_2N\delta_2K\r) +\mathcal{O}(5)\,,\nn\\
&&L_{N\mathcal{R}}\delta N\delta\mathcal{R}=L_{N\mathcal{R}}\l(\delta_1N+\f{1}{2}\delta_2N\r)\l(\delta_1\mathcal{R}+\f{1}{2}\delta_2\mathcal{R}\r)\,,\nn\\
&&L_{N\mathcal{U}}\delta N\delta\mathcal{U}=L_{N\mathcal{U}}\l[\delta_1N(-H\delta_1\mathcal{R}-\f{1}{2}H\delta_2 \mathcal{R}+\delta_1 \mathcal{R}^\mu_\nu\delta_1K^\nu_\mu+\f{1}{2}\delta_1 \mathcal{R}^\mu_\nu\delta_2K^\nu_\mu+\f{1}{2}\delta_1 K^\nu_\mu\delta_2\mathcal{R}^\mu_\nu)\r.\nn\\
&&\l.\,\,+\f{1}{2}\delta_2N(-H\delta_1\mathcal{R}-\f{1}{2}H\delta_2 \mathcal{R}+\delta_1 \mathcal{R}^\mu_\nu\delta_1K^\nu_\mu)\r]\,,\nn\\
&&L_{NN\mathcal{S}}(\delta N)^2\delta\mathcal{S}=L_{NN\mathcal{S}}\l[-2H(\delta_1N)^2(\delta_1K+\f{1}{2}\delta_2K)+\delta_1K^\nu_\mu\delta_1K^\mu_\nu(\delta_1N)^2-2H\delta_1N\delta_2N\delta_1K\r]+\mathcal{O}(5)\,,\nn\\
&&L_{NNK}(\delta N)^2\delta K=L_{NNK}\l[(\delta_1N)^2(\delta_1K+\f{1}{2}\delta_2K)+\delta_1N\delta_2N\delta_1K\r]+\mathcal{O}(5)\,,\nn\\
&&L_{NN\mathcal{U}}(\delta N)^2\delta\mathcal{U}=L_{NN\mathcal{U}}\l[(\delta_1N)^2(-H\delta_1\mathcal{R}-\f{1}{2}H\delta_2\mathcal{R}+\delta_1\mathcal{R}^\mu_\nu\delta_1K^\nu_\mu)-H\delta_1N\delta_2N\delta_1\mathcal{R}\r]\,,\nn\\
&&L_{NN\mathcal{R}}(\delta N)^2\delta\mathcal{R}=L_{NN\mathcal{R}}\l[(\delta_1N)^2(\delta_1\mathcal{R}+\f{1}{2}\delta_2\mathcal{R})+\delta_1N\delta_2N\delta_1\mathcal{R}\r]+\mathcal{O}(5)\,,\nn\\
&&L_{NKK}\delta N(\delta K)^2=L_{NKK}\l[(\delta_1K)^2\l(\delta_1N+\f{1}{2}\delta_2N\r)+\delta_1K\delta_2K\delta_1N\r]+\mathcal{O}(5)\,,\nn\\
&&L_\mathcal{S}\delta \mathcal{S}=-6H^2L_\mathcal{S} -2(H\dot{L}_\mathcal{S}+\dot{H}L_{\mathcal{S}})\l(1-\delta_1N+(\delta_1N)^2-(\delta_1N)^3-\f{1}{2}\delta_2N+\f{1}{4}(\delta_2N)^2+\delta_1N\delta_2N-\f{3}{2}(\delta_1N)^2\delta_2N\r)\nn\\
&&\,\,+L_{\mathcal{S}}(\delta_1 K^\nu_\mu \delta_1 K^\mu_\nu+ \delta_1 K^\nu_\mu \delta_2 K^\mu_\nu+\f{1}{4} \delta_2 K^\nu_\mu \delta_2 K^\mu_\nu )\,,\nn\\
&&L_{K\mathcal{S}}\delta\mathcal{S}\delta K=L_{K\mathcal{S}}\l(-2H((\delta_1K)^2+\delta_1K\delta_2K+\f{1}{4}(\delta_2K)^2)+\delta_1K^\mu_\nu\delta_1K^\nu_\mu\delta_1K+\f{1}{2}\delta_1K^\mu_\nu\delta_1K^\nu_\mu\delta_2K+\delta_1K^\mu_\nu\delta_2K^\nu_\mu\delta_1K\r)\,,\nn\\
&&L_K\delta K=3HL_K+\dot{L}_K\l(1-\delta_1N+(\delta_1N)^2-(\delta_1N)^3-\f{1}{2}\delta_2N+\f{1}{4}(\delta_2N)^2+\delta_1N\delta_2N-\f{3}{2}(\delta_1N)^2\delta_2N\r)+\mathcal{O}(5)\,,\nn \\
&&\f{1}{2}L_{KK}(\delta K)^2=\f{1}{2}L_{KK}\l((\delta_1 K)^2+\f{1}{4}(\delta_2 K)^2+\delta_1K\delta_2K\r)\,,\nn\\
&&\f{1}{6}L_{KKK}(\delta K)^3=\f{1}{6}L_{KKK}\l((\delta_1K)^3+\f{3}{2}(\delta_1K)^2\delta_2K\r)\,,\nn\\
&&L_{K\mathcal{R}}\delta \mathcal{R}\delta K=L_{K\mathcal{R}}\l[\delta_1K\delta_1\mathcal{R}+\f{1}{2}\delta_1K\delta_2\mathcal{R}+\f{1}{2}\delta_2K\delta_1\mathcal{R}+\f{1}{4}\delta_2K\delta_2\mathcal{R}\r]\,,\nn\\
&&L_\mathcal{R}\delta \mathcal{R}=L_\mathcal{R}(\delta_1\mathcal{R}+\f{1}{2}\delta_2\mathcal{R})\,,\nn\\
&&L_\mathcal{U}\delta \mathcal{U}=\f{1}{2}\l[L_{\mathcal{U}}(-3H(\delta_1\mathcal{R}+\f{1}{2}\delta_2\mathcal{R})+\delta_1\mathcal{R}\delta_1K+\f{1}{2}\delta_1\mathcal{R}\delta_2K+\f{1}{2}\delta_2\mathcal{R}\delta_1K+\f{1}{4}\delta_2K\delta_2\mathcal{R})-\dot{L}_{\mathcal{U}}(\delta_1\mathcal{R}+\f{1}{2}\delta_2\mathcal{R}\r. \nn\\ &&\l.\,\,\,-\delta_1\mathcal{R}\delta_1N-\f{1}{2}\delta_1\mathcal{R}\delta_2N-\f{1}{2}\delta_2\mathcal{R}\delta_1N-\f{1}{4}\delta_2\mathcal{R}\delta_2N+\delta_1\mathcal{R}(\delta_1N)^2+\delta_1\mathcal{R}\delta_1N\delta_2N+\f{1}{2}\delta_2\mathcal{R}(\delta_1N)^2-\delta_1\mathcal{R}(\delta_1N)^3)\r]\,,\nn  \\
&&L_{\tilde{\mathcal{K}}}\delta \tilde{\mathcal{K}}=9H^3L_{\tilde{\mathcal{K}}}+3(2H\dot{H}L_{\tilde{\mathcal{K}}}+H^2\dot{L}_{\tilde{\mathcal{K}}})\l(1-\delta_1N+(\delta_1N)^2-(\delta_1N)^3-\f{1}{2}\delta_2N+\f{1}{4}(\delta_2N)^2+\delta_1N\delta_2N-\f{3}{2}(\delta_1N)^2\delta_2N\r) \nn\\
&&+L_{\tilde{\mathcal{K}}}\l(-3H(\delta_1K^\mu_\nu\delta_1K^\nu_\mu+\delta_1K^\mu_\nu\delta_2K^\nu_\mu+\f{1}{4}\delta_2K^\mu_\nu\delta_2K^\nu_\mu) +\delta_1K_{i}^j\delta_1K^{i}_k\delta_1K_{j}^k+\f{3}{2}\delta_1K_{i}^j\delta_1K^{i}_k\delta_2K_{j}^k\r)+\mathcal{O}(5)\,,\nn\\
&&L_{N\tilde{\mathcal{K}}}\delta \tilde{\mathcal{K}}\delta N=  L_{N\tilde{\mathcal{K}}}\l[\delta_1N\l(3H^2\delta_1K+\f{3}{2}H^2\delta_2 K -3H(\delta_1K^\mu_\nu\delta_1K^\nu_\mu+\delta_1K^\mu_\nu\delta_2K^\nu_\mu)\r)+\f{1}{2}\delta_2N(3H^2\delta_1K+\f{3}{2}H^2\delta_2 K \r.\nn\\&&\,\,\,\l.-3H\delta_1K^\mu_\nu\delta_1K^\nu_\mu)+\delta_1 N \delta_1 K_{ij}\delta_1 K^{ik}\delta_1 K^j_k\r] \,,\nn \\
&&L_{NN\tilde{\mathcal{K}}}\delta \tilde{\mathcal{K}}(\delta N)^2=L_{NN\tilde{\mathcal{K}}}\l[(\delta_1N)^2(3H^2\delta_1K+\f{3}{2}H^2\delta_2K-3H\delta_1K^\mu_\nu\delta_1K^\mu_\nu)+3H^2\delta_1N\delta_2N\delta_1K\r]\,,\nn\\
&&L_{\alpha_1}\delta_{\alpha_1}=L_{\alpha_1}\l(\delta_1\alpha_1+\f{1}{2}\delta_2\alpha_1\r)\,,\nn\\
&&L_{\alpha_1N}\delta \alpha_1\delta N= L_{\alpha_1N}\l(\delta_1\delta N\delta_1\alpha_1+\f{1}{2}\delta_1N\delta_2\alpha_1+\f{1}{2}\delta_2N\delta_1\alpha_1\r)+\mathcal{O}(5)\,,\nn\\
&&L_{NK\mathcal{S}}\delta N\delta K \delta \mathcal{S}= -L_{NK\mathcal{S}}H\l(2\delta_1N\delta_1K\delta_2K+\delta_2N\delta_1K^2\r)+\mathcal{O}(5)\,,\nn\\
&&L_{NK\mathcal{R}}\delta N\delta K \delta \mathcal{R}= \f{1}{2}L_{NK\mathcal{R}}  \l( \delta_1N\delta_1K\delta_2\mathcal{R}+\delta_2N\delta_1K\delta_1\mathcal{R} +\delta_1N\delta_2K\delta_1\mathcal{R}\r)+\mathcal{O}(5)\,,
\ea
where we have simplified some of the terms by using a number of useful identities already derived in the literature \cite{Gleyzes:2013ooa}
\ba
\int{}d^4x\sqrt{-g}B(t)K&=&\int{}d^4x\sqrt{-g}\f{\dot{B}}{N}=\int{}d^4x\sqrt{-g}\dot{B}\l(1-\delta_1N+(\delta_1N)^2-(\delta_1N)^3-\f{1}{2}\delta_2N+\f{1}{4}(\delta_2N)^2\r.\nn\\
&+&\l.\delta_1N\delta_2N-\f{3}{2}(\delta_1N)^2\delta_2N\r)+\mathcal{O}(5)\,,
\ea
and
\ba
\int d^4x\sqrt{g}\lambda(t) \mathcal{R}_{\mu\nu}K^{\mu\nu}&=&\int d^4 x\sqrt{g}\l(\frac{\lambda(t)}{2}\mathcal{R}K-\frac{\dot{\lambda}(t)}{2N}\mathcal{R}\r).
\ea

\section{The Kernels for the quartic action}\label{App:Kernels}

In this Appendix we list the Kernels used to define action (\ref{action12final}). They are:

\begin{eqnarray}
K_1(k,k_1)&=&\f{1}{24}\l(   -12\mathcal{W}_0\f{k^2+k_1^2-k\cdot k_1}{a^2} +6\mathcal{W}_5\mathcal{W}_6^2 \f{k^2 k_1^2+2(k\cdot k_1 )(k-k_1)\cdot k_1-(k\cdot k_1)(-2k\cdot k_1 +k\cdot(2k+k_1))}{a^4 \mathcal{W}_4^2}\r.\nn\\
&&\l.+\f{3\mathcal{W}_5\mathcal{W}_6^2}{\mathcal{W}_4^2}\f{k^2k_1^2-16k_1^2k\cdot k_1+3(k\cdot k_1)^2}{a^4}+\f{36M_1\mathcal{W}_6^3}{\mathcal{W}_4^3 a^6}\big( k^2 k_1^2 (k-k_1)^2-2k_1^2 [k\cdot (k-k_1)]^2-k^2[k_1\cdot (k-k_1)]^2\r.\nn\\
&&\l.+2(k\cdot k_1)[k\cdot(k-k_1)]\{k_1\cdot(k-k_1) \}\big)
      \r),\nn\\
K_2(k,k_1)&=&\f{1}{24} \Big[\f{6\mathcal{W}_5\mathcal{W}_6(3\mathcal{W}_4^2+2\mathcal{W}_1\mathcal{W}_5)}{a^2\mathcal{W}_4^3}\{k^2+2\f{(k\cdot k_1)k_1\cdot(k-k_1)}{k_1^2}-\f{k\cdot k_1}{k_1^2}(-2k\cdot k_1+k\cdot(2k+k_1))  \}  \nn\\
&&  \f{3\mathcal{W}_5\mathcal{W}_6(3\mathcal{W}_4^2+2\mathcal{W}_1\mathcal{W}_5)}{a^2\mathcal{W}_4^3}\{k^2-16k\cdot k_1+3\f{(k\cdot k_1)^2}{k_1^2}  \}- \f{36M_1\mathcal{W}_6^2(3\mathcal{W}_4^2+2\mathcal{W}_1\mathcal{W}_5)}{a^4\mathcal{W}_4^4} \{ k^2(k-k_1)^2\nn\\
&&-2 (k\cdot(k-k_1))^2-\f{k^2(k_1\cdot(k-k_1))}{k_1^2}+2\f{(k\cdot k_1)[k\cdot(k-k_1)]\{k_1\cdot(k-k_1) \}}{k_1^2} \}\Big],\nn\\ 
K_3(k,k_1)&=&\Big[  \f{\mathcal{W}_5\mathcal{W}_6}{\mathcal{W}_4^2}(-8\f{k^2k_1^2M_4}{a^4}+8\f{(k\cdot k_1)^2M_4}{a^4} -\f{\mathcal{W}_4}{2a^2}[k^2-k_1^2+2k\cdot k_1] ) 
+\f{\mathcal{W}_6\mathcal{W}_5}{2a^2\mathcal{W}_4}(-k_1^2+k\cdot k_1)\nn\\
&&+\f{\mathcal{W}_5\mathcal{W}_6(3\mathcal{W}_4^2+2\mathcal{W}_1\mathcal{W}_5)}{8a^4 \mathcal{W}_4^4(k-k_1)^2}(k^2 k_1^2-16k_1^2(k\cdot k_1)+3(k\cdot k_1)^2)+\f{\mathcal{W}_6^2(3\mathcal{W}_4^2+2\mathcal{W}_1\mathcal{W}_5)}{a^4\mathcal{W}_4^4}\big( \f{3M_1k^2 (k_1)^2}{2}\nn\\
&&-3\f{k_1^2}{(k-k_1)^2}(k\cdot\{k-k_1\})^2M_1-3M_1\f{k^2(k_1\cdot\{k-k_1\})^2+2[k\cdot k_1][k_1\cdot (k-k_1)][k\cdot (k-k_1)]}{2(k-k_1)^2}\big)\nn\\
&&+\f{\mathcal{W}_5\mathcal{W}_6}{2a^2\mathcal{W}_4}(2k^2+k_1^2-2k\cdot k_1)+\f{\mathcal{W}_6^2 M_1}{a^4\mathcal{W}_4^2}(3k^2k_1^2+3(k\cdot k_1)^2)
+\f{\mathcal{W}_5 \mathcal{W}_6^2}{a^4\mathcal{W}_4^3}(X_{10}(k\cdot k_1)^2+X_4 k^2 k_1^2)\Big],\nn\\
K_4(k,k_1)&=&\f{1}{24}\Big[ \f{36\mathcal{W}_1\mathcal{W}_5^2}{\mathcal{W}_4^2}+\f{12\mathcal{W}_5(3\mathcal{W}_4^2+2\mathcal{W}_1\mathcal{W}_5)}{a^2 \mathcal{W}_4^3}\big[ M_4(-8k^2+8\f{(k\cdot k_1)^2}{k_1^2})+\mathcal{W}_4 a^2(1-\f{k\cdot k_1}{k_1^2})\big]\nn\\
 &&+\f{3\mathcal{W}_5(3\mathcal{W}_4^2+2\mathcal{W}_1\mathcal{W}_5)^2}{a^2 (k-k_1)^2\mathcal{W}_4^4}\big(k^2-16k\cdot k_1+3\f{(k\cdot k_1)^2}{k_1^2} \big)+\f{12\mathcal{W}_5}{a^2 \mathcal{W}_4^2}\big(16k^2M_4 \mathcal{W}_4+a^2(2(6\mathcal{W}_4^2+\mathcal{W}_1\mathcal{W}_5)\nn\\
 &&-\f{(k\cdot k_1)}{k_1^2}(3\mathcal{W}_4^2+2\mathcal{W}_1\mathcal{W}_5)) \big)+\f{256k^2M_1\mathcal{W}_6}{a^2 \mathcal{W}_4}-\f{36(3\mathcal{W}_4^2+2\mathcal{W}_1\mathcal{W}_5)^2}{k_1^2 (k-k_1)^2a^2\mathcal{W}_4^5}\big\{k^2k_1^2(k-k_1)^2-2k_1^2(k\cdot(k-k_1))^2\nn\\
 &&-k^2 (k_1\cdot (k-k_1))^2+2(k\cdot k_1)[k\cdot(k-k_1)][k_1\cdot(k-k_1)] \big\} +\f{24k^2\mathcal{W}_5^2\mathcal{W}_6X_2}{a^2\mathcal{W}_4^3}+\f{24k^2\mathcal{W}_5^2X_3}{a^2 \mathcal{W}_4^2}\nn\\
 &&\f{24(3\mathcal{W}_4^2+2\mathcal{W}_1\mathcal{W}_5)\mathcal{W}_6\mathcal{W}_5}{a^2\mathcal{W}_4}(k^2X_4+X_{10}\f{(k\cdot k_1)^2}{k_1^2})+\f{24\mathcal{W}_6}{a^2\mathcal{W}_4^3}\{ 3(k^2+\f{(k\cdot k_1)^2}{k_1^2})M_1(3\mathcal{W}_4^2+2\mathcal{W}_1\mathcal{W}_5)\nn\\
 &&-k^2X_8 \mathcal{W}_5\mathcal{W}_4 \}\Big]\,,\nn\\
 K_{1d}(k,k_1)&=&\Big[-\f{\mathcal{W}_5\mathcal{W}_6(k_1^2+k\cdot k_1)}{4a^2\mathcal{W}_4}+\f{\mathcal{W}_5\mathcal{W}_6}{\mathcal{W}_4^2}(\f{-4M_4k_1^2(k-k_1)^2+4M_4(k_\cdot(k-k_1))^2}{a^4}-\mathcal{W}_4\f{k_1^2-k_1\cdot(k-k_1)}{2a^2})\nn\\
 &&+\f{\mathcal{W}_5\mathcal{W}_6}{2a^2\mathcal{W}_4}(k_1^2+k_1\cdot(k-k_1))  +\f{(3\mathcal{W}_4^2+2\mathcal{W}_1\mathcal{W}_5)\mathcal{W}_6\mathcal{W}_5}{4a^4\mathcal{W}_4^3}(k_1^2-2\f{(k\cdot k_1)(k\cdot(k-k_1))}{k^2}-\f{(k\cdot k_1)^2}{k^2})\nn\\
 &&+2\f{(k\cdot k_1)(k_1\cdot(k-k_1))}{k^2}+\f{\mathcal{W}_6^2M_1}{2a^4\mathcal{W}_4^2}(3k_1^2(k-k_1)^2-12(k_1\cdot(k-k_1))^2)\nn\\
&&-\f{\mathcal{W}_6^2(3\mathcal{W}_4^2+2\mathcal{W}_1\mathcal{W}_5)M_1}{2a^4\mathcal{W}_4^4}(3k_1^2(k-k_1)^2-6\f{k_1^2(k\cdot(k-k_1))^2}{k^2}-3(k_1\cdot(k-k_1))^2\nn\\
 &&+6\f{[k\cdot k_1][k\cdot(k-k_1)][k_1\cdot(k-k_1)]}{k^2}) -\f{\mathcal{W}_5\mathcal{W}_6^2}{2a^4\mathcal{W}_4^3}(X_{10}(k_1\cdot(k-k_1))^2+X_4k_1^2(k-k_1)^2)\Big]\,,\nn\\
 K_{2d}(k,k_1)&=&\Big[ \f{\mathcal{W}_5(3\mathcal{W}_4^2+2\mathcal{W}_1\mathcal{W}_5)}{a^2\mathcal{W}_4^3}(-4M_4k_1^2(k-k_1^2)+4\f{(k_1\cdot(k-k_1))^2M_4}{k_1^2}-\mathcal{W}_4a^2(1/2-\f{k\cdot(k-k_1)}{2k_1^2})) \nn\\
 &&    - \f{\mathcal{W}_5(3\mathcal{W}_4^2+2\mathcal{W}_1\mathcal{W}_5)}{\mathcal{W}_4^2}  (-1/2 -\f{k_1\cdot (k-k_1)}{k_1^2}) +\f{\mathcal{W}_5(3\mathcal{W}_4^2+2\mathcal{W}_1\mathcal{W}_5)^2}{4\mathcal{W}_4^4}(1-2\f{k_1\cdot(k-k_1)}{k_1^2}-\f{(k\cdot k_1)^2}{k^2 k_1^2}\nn\\
&&+2\f{(k\cdot k_1)(k_1\cdot (k-k_1))}{k^2 k_1^2}) +\f{(3\mathcal{W}_4^2+2\mathcal{W}_1\mathcal{W}_5)\mathcal{W}_6 M_1}{2a^2\mathcal{W}_4^3}(3(k-k_1)^2-12\f{(k_1\cdot(k-k_1))^2}{k_1^2})\nn\\
&&-\f{\mathcal{W}_6 M_1(3\mathcal{W}_4^2+2\mathcal{W}_1\mathcal{W}_5)^2}{2a^2\mathcal{W}_4^5}(3(k-k_1)^2-6\f{(k\cdot(k-k_1))^2}{k^2}-3\f{(k_1\cdot(k-k_1))^2}{k_1^2}\nn\\
&&+6\f{(k\cdot k_1)(k\cdot (k-k_1))(k_1\cdot(k-k_1))}{k^2k_1^2}) -\f{\mathcal{W}_6 \mathcal{W}_5(3\mathcal{W}_4^2+2\mathcal{W}_1\mathcal{W}_5)}{2a^2\mathcal{W}_4^4}(X_{10}\f{(k_1\cdot(k-k_1))^2}{k_1^2}+X_4(k-k_1)^2) \Big],\nn\\
K_{3d}(k,k_1)&=&\Big[ -\f{9\mathcal{W}_5}{2}+\f{\mathcal{W}_5}{2a^2\mathcal{W}_4}(16M_4k_1^2+9a^2\mathcal{W}_4)+\f{\mathcal{W}_5(3\mathcal{W}_4^2+2\mathcal{W}_1\mathcal{W}_5)}{2a^2\mathcal{W}_4^3}  (-8k_1^2M_4+8\f{(k\cdot k_1)^2}{k^2}M_4-a^2\mathcal{W}_4\nn\\
&&-\f{k\cdot k_1}{k^2}a^2\mathcal{W}_4)+\f{54k_1^2M_1\mathcal{W}_6}{a^2\mathcal{W}_4}+\f{\mathcal{W}_6 M_1(3\mathcal{W}_4^2+2\mathcal{W}_1\mathcal{W}_5)}{2a^4\mathcal{W}_4^3} (3k_1^2-12\f{(k_1\cdot(k-k_1))^2}{(k-k_1)^2}) \nn\\
&&-\f{\mathcal{W}_6 M_1(3\mathcal{W}_4^2+2\mathcal{W}_1\mathcal{W}_5)^2}{2a^2\mathcal{W}_4^5}(3k_1^2-6\f{(k\cdot(k-k_1))^2}{(k-k_1)^2}-3\f{(k_1\cdot(k-k_1))^2}{(k-k_1)^2}\nn\\&& +6\f{[k\cdot k_1][k\cdot(k-k_1)][k_1\cdot(k-k_1)]}{k^2(k-k_1)^2})
-\f{(3\mathcal{W}_4^2+2\mathcal{W}_1\mathcal{W}_5)}{2\mathcal{W}_4^2}(\mathcal{W}_5\f{k\cdot k_1}{k^2}\nn\\
&&-\mathcal{W}_5-\f{2M_1\mathcal{W}_6}{a^2\mathcal{W}_4}(3k_1^2+\f{(k\cdot k_1)^2}{k^2})) -\f{k_1^2\mathcal{W}_5^2\mathcal{W}_6 X_2}{2a^2\mathcal{W}_4^3} +\f{\mathcal{W}_5^2}{\mathcal{W}_4^2}(3\mathcal{W}_1+\f{2k_1^2X_3}{a^2})\nn\\
&&-    \f{(3\mathcal{W}_4^2+2\mathcal{W}_1\mathcal{W}_5)\mathcal{W}_6\mathcal{W}_5}{2\mathcal{W}_4^4 a^2}(X_{10}(2\f{(k\cdot k_1^2)^2}{k^2}+\f{(k_1\cdot (k-k_1))^2}{(k-k_1)^2})+X_4(3 k_1^2)-\f{k_1^2\mathcal{W}_5\mathcal{W}_6 X_8}{a^2\mathcal{W}_4^2}\nn\\
&&-\f{\mathcal{W}_5(k^2\mathcal{W}_6X_8/\mathcal{W}_4-9\mathcal{W}_4a^2/2-8k_1^2 M_4)}{a^2\mathcal{W}_4} \Big]\,,\nn\\
K_{4d}(k,k_1)&=& \Big[-36M_1+\f{194M_1(3\mathcal{W}_4^2+2\mathcal{W}_1\mathcal{W}_5)}{3\mathcal{W}_4^2} +\f{M_1(3\mathcal{W}_4^2+2\mathcal{W}_1\mathcal{W}_5)^2}{\mathcal{W}_4^4}(+3/2-6\f{(k_1\cdot (k-k_1))^2}{k_1^2(k-k_1)^2}) \nn\\
&& -\f{M_1(3\mathcal{W}_4^2+2\mathcal{W}_1\mathcal{W}_5)^3}{\mathcal{W}_4^6}(3/2-3\f{(k_1\cdot (k-k_1))^2}{k_1^2(k-k_1)^2}-3\f{(k\cdot(k-k_1))^2}{k^2(k-k_1)^2}+3\f{[k\cdot k_1][k\cdot(k-k_1)][k_1\cdot(k-k_1]}{k^2k_1^2(k-k_1)^2}) \nn\\
&&  -\f{X_1 \mathcal{W}_5^3}{\mathcal{W}_4^3}+2\f{X_2\mathcal{W}_5^2}{\mathcal{W}_4^2}-3\f{X_2(3\mathcal{W}_4^2+2\mathcal{W}_1\mathcal{W}_5)\mathcal{W}_5^2}{2\mathcal{W}_4^4}-\f{M_1(3\mathcal{W}_4^2+2\mathcal{W}_1\mathcal{W}_5)^2}{\mathcal{W}_4^5}(\f{X_4}{2}+X_{10}\f{(k_1\cdot (k-k_1))^2}{2k_1^2(k-k_1)^2})\nn\\
&& +\f{9\mathcal{W}_5 X_8}{2\mathcal{W}_4}-\f{X_8(3\mathcal{W}_4^2+2\mathcal{W}_1\mathcal{W}_5)\mathcal{W}_5}{\mathcal{W}_4^3}-\f{(3\mathcal{W}_4^2+2\mathcal{W}_1\mathcal{W}_5)}{\mathcal{W}_4^2} \big\{\f{M_1(3\mathcal{W}_4^2+2\mathcal{W}_1\mathcal{W}_5)^3}{\mathcal{W}_4}(-3-3\f{(k\cdot k_1)^2}{k^2k_1^2}) +\f{X_8 \mathcal{W}_5}{\mathcal{W}_4}  \big\}\nn\\
&&-\f{\mathcal{W}_5}{\mathcal{W}_4}(\f{k^2X_8(3\mathcal{W}_4^2+2\mathcal{W}_1\mathcal{W}_5)}{k_1^2\mathcal{W}_4^2}-\f{6\mathcal{W}_5X_2}{\mathcal{W}_4})\Big]\,.
\end{eqnarray}

\end{document}